\documentclass[aps,pra,groupedaddress,preprint]{revtex4}
\usepackage{amsmath}
\usepackage{amssymb}
\usepackage{graphics}
\usepackage{graphicx}
\usepackage{color}

\newcommand \beq{\begin{eqnarray}}
\newcommand \eeq{\end{eqnarray}}
\def\simge{\mathrel{%
       \rlap{\raise 0.511ex \hbox{$>$}}{\lower 0.511ex \hbox{$\sim$}}}}
\def\simle{\mathrel{
       \rlap{\raise 0.511ex \hbox{$<$}}{\lower 0.511ex \hbox{$\sim$}}}}

\usepackage{graphicx,color}
\usepackage{amsmath,amssymb}
\usepackage{feynmf}
\newcommand{\Slash}[1]{{\ooalign{\hfil/\hfil\crcr$#1$}}}

\newcommand{\vp}{\vec{p}}

\newcommand{\la}{\langle}
\newcommand{\ra}{\rangle}

\newcommand{\calL}{\mathcal{L}}

\newcommand{\calE}{\mathcal{E}}
\newcommand{\calA}{\mathcal{A}}

\newcommand{\calD}{\mathcal{D}}
\newcommand{\calN}{\mathcal{N}}
\newcommand{\calM}{\mathcal{M}}

\newcommand{\calF}{\mathcal{F}}

\newcommand{\rmd}{\mathrm{d}}
\newcommand{\rmi}{\mathrm{i}}
\newcommand{\rme}{\mathrm{e}}

\newcommand{\tp}{ \tilde{p} }

\newcommand{\tQ}{ \tilde{Q} }

\newcommand{\tomega}{ \tilde{\omega} }
\newcommand{\tvarphi}{ \tilde{\varphi} }
\newcommand{\tchi}{ \tilde{\chi} }

\newcommand{\sn}{\mathrm{sn}}
\newcommand{\cn}{\mathrm{cn}}
\newcommand{\dn}{\mathrm{dn}}
\newcommand{\bfK}{{\bf K}}
\newcommand{\bfE}{{\bf E}}

\begin{document}
\title{Chiral spirals from noncontinuous chiral symmetry:\\
The Gross-Neveu model results}
\author{Toru Kojo}
\affiliation{Department of Physics, University of Illinois, 1110
  West Green Street, Urbana, Illinois 61801, USA}  
\date{\today}

\begin{abstract}
It is shown that
the inhomogeneous chiral condensate
in the Gross-Neveu (GN) model 
takes the chiral spiral form,
even though the thermodynamic functional
depends only on the chiral scalar density.
It is the inhomogeneity
of the chiral scalar condensate
that drives the spatial modulations of the
pseudoscalar one.
The result has broader implications
once we start to think of fundamental theories
behind the effective models.
In particular, some effective interactions---which 
may be omitted for descriptions of the homogeneous phases---can 
be dynamically enhanced due to the spatial modulations
of the large mean fields.
Implications for the four-dimensional counterparts
of the GN model are discussed.
In a quark matter context, 
proper forms of the effective models 
for the inhomogeneous phases
are speculated, through considerations on 
the Fermi-Dirac sea coupling.
\end{abstract}

\maketitle

\section{Introduction}

Recently, phases of the 
inhomogeneous chiral condensates (IChCs)
have attracted renewed attention in the quark matter 
context \cite{Deryagin:1992rw,Buballa:2014tba}.
A number of studies based on
the NJL-type model \cite{Nickel:2009ke,Abuki:2011pf}
as well as models
with the infrared (IR) 
enhanced interactions \cite{Kojo:2009ha,Kojo:2011cn}
have suggested that in some domain of
moderate quark density
the IChC phases
are energetically more favored 
than the normal, chiral symmetric phase.
In particular, the NJL-type model studies indicate that
the phase of IChC may mask
the usual first-order chiral 
phase transition line and its critical end point,
and might change the conventional wisdom.

So far, most studies have been concentrated on 
the chiral condensates of the liquid crystal type,
in which the condensates spatially modulate 
in one particular direction (say, the $z$ direction),
while they are uniform in the other two directions.
For the description of such phases, 
the model studies
rely on the understanding of their two-dimensional counterparts:
the Gross-Neveu (GN) model \cite{Thies:2003kk,Thies:2003br}
as a counterpart for the NJL$_4$ model \cite{Nickel:2009ke},
the 't Hooft model (QCD$_2$) \cite{Schon:2000he}
for the confining model \cite{Kojo:2009ha,Kojo:2011cn},
and the NJL$_2$ model \cite{Basar:2009fg}
for the extended NJL$_4$ model with
tensor 4-Fermi interactions \cite{Feng:2013tqa}.
In fact, the solutions of two-dimensional models
can be naturally embedded into 
the four-dimensional mean field ansatz.

The inhomogeneous solutions for two-dimensional models
are similar but not quite identical.
The QCD$_2$ and NJL$_2$ models are known to 
have the chiral spiral ground states,
\begin{equation}
\la \bar{\psi} \psi \ra_{ {\rm 2D} }  
= \Delta \cos \left( 2p_F z \right)\,,
~~~~~~
\la \bar{\psi} \, \rmi \gamma_0 \gamma_z \psi \ra_{ {\rm 2D} }  
= \Delta \sin \left( 2p_F z \right) \,,
~~~~(\gamma^{{\rm 2D}}_5 = \gamma_0 \gamma_z)
\end{equation}
which can be directly brought into its four-dimensional 
version by replacing 
$\left(\gamma_0 \gamma_z\right)^{ {\rm 2D} } 
\rightarrow \left(\gamma_0 \gamma_z\right)^{ {\rm 4D} } $.
(This solution should {\it not} be confused with
the pionic chirals such as
$\la \bar{\psi} \, \rmi \gamma_5 \psi \ra_{ {\rm 4D} } $
or $\la \bar{\psi} \, \rmi \tau_3 \gamma_5 \psi \ra_{ {\rm 4D} } $
where $\gamma_5^{ {\rm 4D} } \neq 
\left(\gamma_0 \gamma_z\right)^{ {\rm 4D} }$ \footnote{The 
chiral spirals made of the chiral scalar and tensor
condensates are dominated by pairs of particle-holes
near the Fermi surface, and produce gaps.
In contrast, the pionic chiral spirals in addition contain
particle-antiparticle pairing,
and have gaps in both the Fermi and Dirac seas.
Discussions in Sec. \ref{sec:discussions}
can be used to understand differences between
these two types of chiral spirals.
}.)
Here $2p_F$ appears because of the condensed pairs of
comoving particle-holes
near the Fermi surface. 
On the other hand, for the GN model,
the spiral solution is usually not considered,
because the 4-Fermi interaction takes the form
$(\bar{\psi} \psi)^2$,
so that its mean-field thermodynamic functional
depends only on $\la \bar{\psi} \psi \ra$
but not on $\la \bar{\psi} \rmi \gamma_5 \psi \ra$.
Therefore, the above two classes of solutions
are usually distinguished.

In this paper we explain how to understand differences 
between them
by revisiting inhomogeneous solutions 
of the GN model \cite{Thies:2003br}.
To avoid confusion,
we emphasize that we will not attempt
to modify the analytic solution,
which was shown to
achieve the ground state \cite{Thies:2003kk}.
On the other hand,
there are physical implications
which cannot be observed from the expression of 
the thermodynamic functional.
In fact, not all condensates 
manifestly appear in the energy minimization procedure.

Using the analytically known fermion eigenstates, 
we compute condensates explicitly
to show that the inhomogeneous
condensate in the GN model actually takes 
the chiral spiral form.
It is, however, not identical with those in QCD$_2$ (NJL$_2$).
In the GN model, the net contribution to
the chiral scalar density comes from the Dirac sea, 
while that for the chiral pseudoscalar density comes 
from the Fermi sea.
This introduces disparities in amplitudes of two densities.

This chiral spiral solution in the GN model
can be elevated to the NJL$_4$ model.
Like the GN model case, the corresponding chiral 
spiral--which is made of 
$\la \bar{\psi} \psi \ra_{ {\rm 4D} }$ and
$\la \bar{\psi} \, \rmi \gamma_0 \gamma_z \psi \ra_{ {\rm 4D}
}$--cannot be observed from the thermodynamic functional
in the NJL$_4$ model, and it must be computed
using the fermion bases
in the scalar mean field of Ref. \cite{Nickel:2009ke}.
We will show the explicit mapping 
from two to four dimensions in another publication,
but we think that 
the main features should already be clear from 
our two-dimensional analyses in this paper.

Actually, for the discussions of the QCD phase diagram,
the derivation of the chiral spirals in the NJL$_4$ model
is not the end of the story. 
It leads to broader implications 
once we start to think of fundamental theories
behind the effective models.

For the NJL$_4$ model up to dimension-6 operators,
in principle we should include all possible
4-Fermi interactions 
that are compatible with symmetries of QCD,
although many of them can be discarded
based on other set of arguments.
For example, in vacuum, 
it does not matter whether or not we include
tensor type interactions
$\sim \left(\bar{\psi} \sigma_{\mu \nu} \psi\right)^2 
+ \left(\bar{\psi} \rmi \gamma_5 \sigma_{\mu \nu} \tau_a\psi\right)^2$,
simply because the tensor mean field is zero,
not because the coupling constant is small
(there are no reasons why the coupling should be very small).
The only important mean field comes from the scalar density,
so that terms
$\sim (\bar{\psi} \psi)^2 + (\bar{\psi} \rmi \gamma_5 \tau_a\psi)^2$
are enough to take into account relevant dynamical effects
and, at the same time, maintain the chiral symmetry.

The situation is different for inhomogeneous phases.
As explained above, 
the spatially modulating scalar density 
drives the spatial modulation of the tensor mean field
$\la \bar{\psi} \rmi \gamma_0 \gamma_z \psi\ra$
whose amplitude is comparable to the scalar one.
In this case, the relevance of the tensor-type interactions
is dynamically enhanced,
so we have to reanalyze the mean field solutions
in the presence of such interactions.
If the new mean fields turn out to generate another
mean field, again we have to include the corresponding 
4-Fermi interactions and reanalyze dynamics
from the beginning.
This procedure should be repeated until 
we exhaust all possible 
dynamically enhanced 4-Fermi interactions
and mean fields.
After that, we can pick out the effective
models for the inhomogeneous phase.

This paper is organized as follows:
In Sec. \ref{sec:GN},
we review the inhomogeneous
mean field solution for the GN model
and reproduce a number of important
results in Ref. \cite{Thies:2003br}.
We quickly summarize basics of the elliptic functions,
to the extent necessary for converting 
the mathematical structure into physical terminology.
In Sec. \ref{sec:condensates},
we calculate the expectation values of various operators--
in particular, pseudoscalar density.
By examining its relationship with the scalar density,
we show that they form the chiral spirals
with unequal amplitudes.
In Sec. \ref{sec:discussions},
we compare the chiral spirals in
the GN model to QCD$_2$ and the NJL$_2$ model.
Section \ref{sec:summary} is devoted to summary.

In Sec. \ref{sec:GN} and the appendixes, we add a number of supplementary 
materials for Ref. \cite{Thies:2003br},
because the descriptions in the original paper 
were rather dense and hard to access for nonexperts.
We try to reduce the gaps between calculations
in Ref. \cite{Thies:2003br}.
The relevant formula to be used can be found in 
a handbook for mathematics \cite{formula1},
and its derivation can be found in Ref. \cite{analysis}.
Throughout this paper,
we use the convention
$(x^0,x^1)=(t,x)$ and $g_{\mu \nu}={\rm diag.}(1,-1)$.

\section{Inhomogeneous mean fields for 
the Gross-Neveu model}
\label{sec:GN}

The Gross-Neveu model with $N$ colors is
\begin{equation}
\calL = \bar{\psi} \, \rmi \Slash{\partial} \psi 
+ \frac{G}{\, 2N \,} \left(\bar{\psi}\psi\right)^2 
\,,
\end{equation}
where the sum over color indices are implicit.
We consider $N\rightarrow \infty$ for the mean field considerations.
Using the auxiliary field method, we have
\begin{equation}
\calL = \bar{\psi} \, 
\left[\, \rmi \Slash{\partial} -M (x) \,\right]\psi 
- \frac{N}{\, 2G \,} M^2 (x) \,.
\end{equation}
We are going to use the canonical approach
to treat the system at finite density.
The constraint will be treated in Sec. \ref{sec:condensates},
while in this section we just investigate
properties of the eigenstates.

In Sec. \ref{sub:fe}, we first review
the mean field Ansatz and some properties
of the elliptic functions.
In Sec. \ref{sub:ee}, we summarize properties
of the fermion eigenstates
such as relations between 
the energy and quasimomentum.
The density of states is given in Sec. \ref{sub:ds}.
How to map the UV cutoff from the homogeneous 
to the inhomogeneous phase
is explained in Sec. \ref{sub:mu}.

\subsection{Field equations}
\label{sub:fe}

\begin{figure}[tb]
\vspace{-0.5cm}
\begin{center}
  \includegraphics[scale=.25]{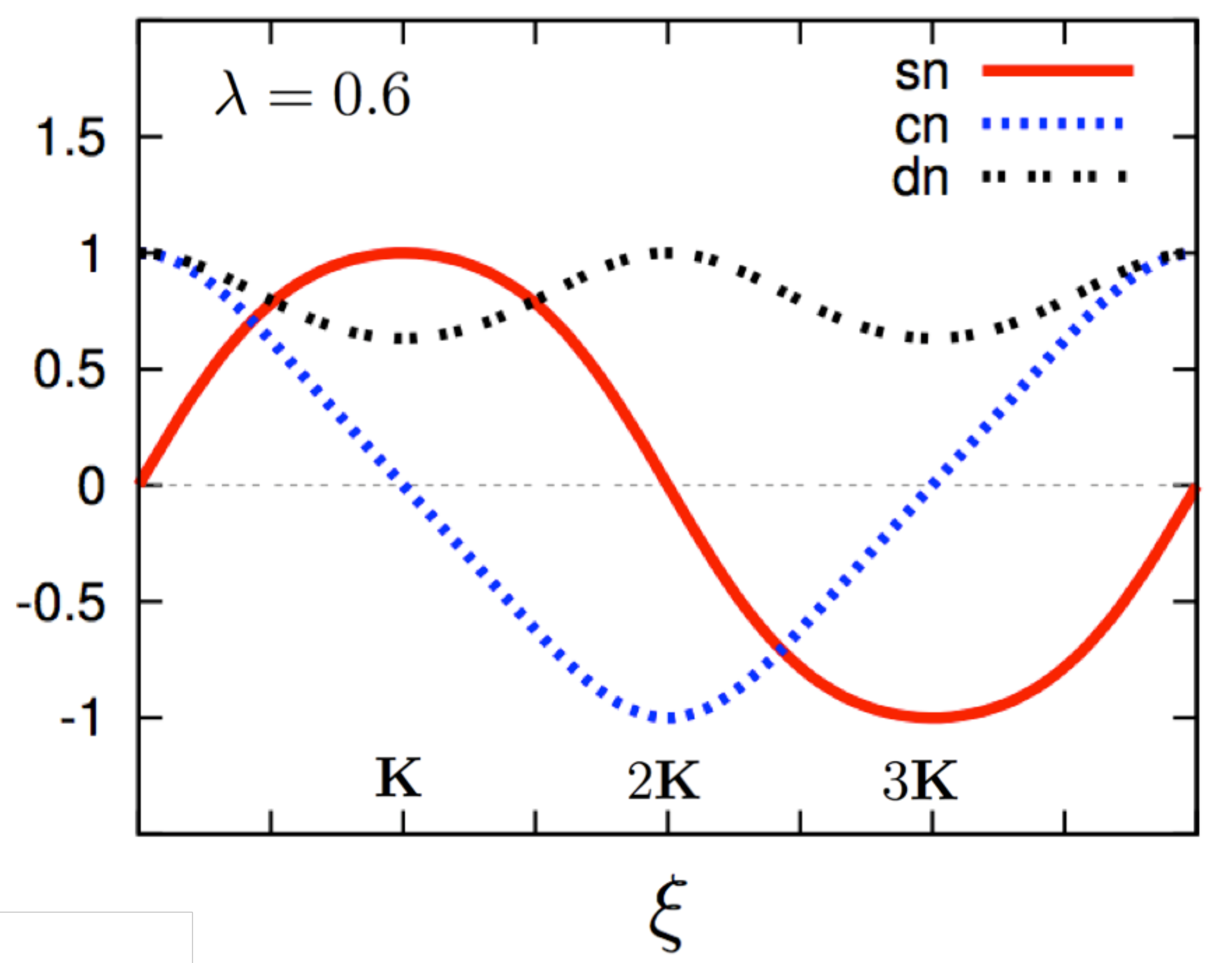} 
\end{center}
\vspace{-0.6cm}
\caption{The Jacobi elliptic functions,
$\sn(\xi|\lambda),\, \cn(\xi|\lambda)$, and
$\dn(\xi|\lambda)$
at $\lambda =0.6$.
$\bfK(\lambda)$ is the quarter period.
}
\label{fig:sncn}
\end{figure}

We first analyze the Dirac equation.
The field equation is
\begin{equation}
\left[-\rmi \gamma^5 \partial_1 + M(x) \gamma^0 \right]\psi
=\omega \psi \,.
\end{equation}
To proceed further,
we choose the $\gamma$ matrices and spinor bases as
\begin{equation}
\gamma^0 =-\sigma_1\,,
~~~~\gamma^1 = \rmi \sigma_3\,,
~~~~\gamma^5= \gamma^0 \gamma^1 
= - \sigma_2\,,
~~~~
\psi_\omega (x) 
= \left[\begin{matrix}
\, \varphi_\omega (x) \, \\
\, \chi_\omega (x) \,
\end{matrix}\right] \,.
\end{equation}
Then the field equation takes the form
\begin{equation}
\left[\, \partial_1 - M(x) \,\right] \chi = \omega \varphi\,,
~~~~~~~
\left[\, -\partial_1 - M(x) \,\right] \varphi = \omega \chi\,.
\label{relation}
\end{equation}
The reason to take the above bases
is that the Dirac equation with a mean field
can be brought into the Lame form,
whose analytic properties have been investigated 
(Refs. \cite{Kusnezov} and \cite{Dunne:1997ia}
are very useful).
From this set of equations,
we can find 
\begin{equation}
\left[-\frac{\partial^2}{\partial x^2}
- \frac{\partial M}{\partial x} + M^2 \right] \varphi
= \omega^2 \varphi\,,
~~~~~~
\left[-\frac{\partial^2}{\partial x^2}
+ \frac{\partial M}{\partial x} + M^2 \right] \chi
= \omega^2 \chi\,.
\label{fieldeq}
\end{equation}
Now we consider the ansatz proposed by Thies 
\cite{Thies:2003br}.
Its form is
\begin{equation}
M(x) = \calA \lambda\,
\frac{\, \sn(\calA x|\lambda) \,\cn(\calA x|\lambda)\,}
{ \dn(\calA x|\lambda) }
\equiv \calA \calM(\xi|\lambda) \,,
~~~~~~ \xi = \calA x\,,
\end{equation}
where $\sn$, $\cn$, and $\dn$ are Jacobi's
elliptic functions with the elliptic modulus $\lambda$.
$\calA$ and $\lambda$ are variational parameters.
To get feelings about the ansatz,
let us briefly look at basic properties
of the elliptic functions:

(i) The elliptic functions interpolate
the trigonometric functions and hyperbolic functions
through the elliptic parameter $\lambda$
\footnote{Formulas (16.13) and (16.15) in Ref. \cite{formula1}.}.
In the $\lambda\rightarrow 0$ limit,
\begin{equation}
\sn(\xi| 0) = \sin \xi\,,
~~~ \cn(\xi|0) = \cos \xi\,,
~~~ \dn(\xi|0) = 1 \,,
\end{equation}
and in the
$\lambda\rightarrow 1$ limit,
\begin{equation}
\sn(\xi| 1) = \tanh \xi\,,
~~~ \cn(\xi|1) = {\rm sech} \, \xi\,,
~~~ \dn(\xi|1) = {\rm sech} \, \xi \,,
\end{equation}
with which 
\begin{equation}
\calM(\xi|\lambda \rightarrow 0) 
\sim \frac{\, \lambda \,}{2} \sin(2\xi) \,,
~~~~~~
\calM(\xi|\lambda \rightarrow 1) 
\sim \pm \tanh (\xi) \,.
\end{equation}
The $\lambda \rightarrow 0$ limit
corresponds to the density wave solution at high density,
and 
the $\lambda \rightarrow 1$ limit
corresponds to solitonic solutions at low density.
As an example, in Fig.\ref{fig:sncn}  
we plot these functions 
for $\lambda=0.6$.
This asymptotic behavior motivates us to
use the ansatz interpolating 
these two solutions
which are known to minimize the thermodynamic potential.

\begin{figure}[tb]
\vspace{-0.5cm}
\begin{center}
  \includegraphics[scale=.25]{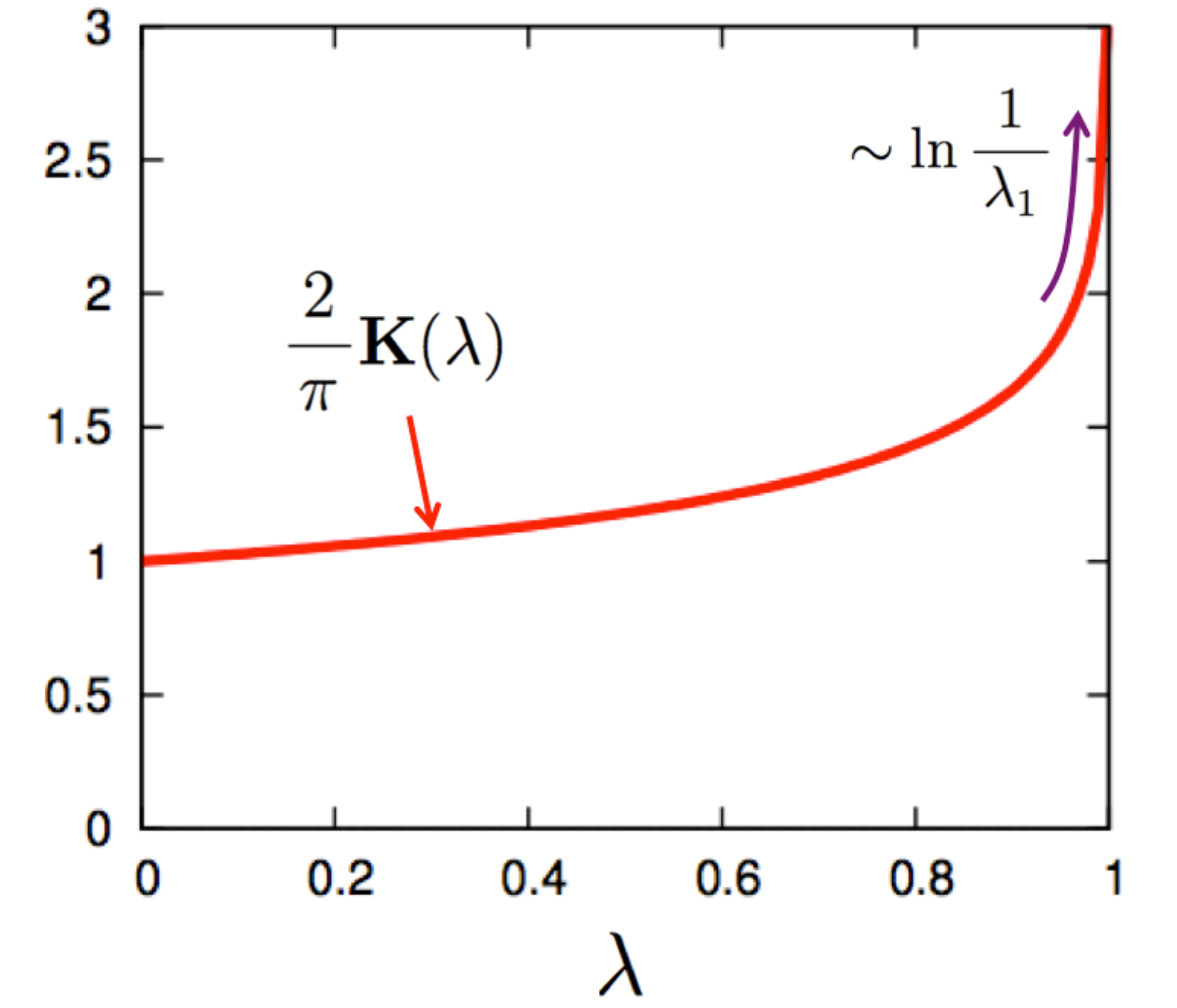} 
\end{center}
\vspace{-0.6cm}
\caption{The Jacobi complete
elliptic function of the first kind $\bfK(\lambda)$
(normalized by $\pi/2$).
The period grows logarithmically as
$\lambda$ approaches $1$,
which corresponds to the dilute limit
of the fermion density.
}
\label{fig:Klambda}
\end{figure}

(ii) As with trigonometric functions,
there are simple square relations,
\begin{equation}
\sn^2(\xi|\lambda ) + \cn^2(\xi|\lambda)  =1\,,
~~~~~~
\dn^2(\xi|\lambda) 
= 1 - \lambda \, \sn^2(\xi|\lambda) \,.
\end{equation}

(iii) As with trigonometric functions, we can define
the quarter period.
It is given by the Jacobi complete 
elliptic integral of the first kind
\footnote{Formulas (17.3.1) and (17.3.26) in Ref. \cite{formula1}.},
\begin{equation}
\bfK (\lambda) 
= \int^{\pi/2}_0 \!
\frac{\rmd \theta}{\, \sqrt{1-\lambda \sin^2 \theta \,} \,}
~~~~~
\rightarrow ~~~\left\{
\begin{matrix}
~~~~~\pi/2~~~~~~~(\lambda \rightarrow 0)\\
~~ \frac{1}{\,2\,} \ln \frac{\, 16\,}{\, \lambda_1 \,}
~~~~~(\lambda \rightarrow 1)
\end{matrix}
\right. \,,
\end{equation}
where $\lambda_1\equiv 1-\lambda$
is called the complementary modulus of $\lambda$.
As the name suggests,
$\sn(\xi+4\bfK)=\sn(\xi)$, 
$\cn(\xi+4\bfK)=\cn(\xi), \cdots$ 
are satisfied for any values of $\lambda$.
The changes for the quarter period
are more nontrivial,
and there are formulas. \footnote{See 
also Table (16.8) in Ref. \cite{formula1}.}
(Below,
we sometimes omit $\lambda$ as long as
it does not bring any confusion.)
\begin{equation}
\sn(\xi\pm\bfK) 
= \pm \frac{\, \cn(\xi) \,}{\, \dn(\xi) \,}\,,
~~~~~
\cn(\xi\pm\bfK ) 
= \mp \lambda_1^{1/2} \frac{\, \sn(\xi) \,}{\, \dn(\xi) \,}\,,
~~~~~
\dn(\xi\pm\bfK) 
= \lambda_1^{1/2} \frac{1}{\, \dn(\xi) \,} \,,
\label{eq:quarter}
\end{equation}
with which we can show
\begin{equation}
\calM(\xi) = - \calM(\xi \pm \bfK) = \calM(\xi \pm 2\bfK)\,.
\label{shift}
\end{equation}
The first equality can be used to cast 
the equation for $\chi$ into the same form as for $\varphi$.
For later convenience, we rescale variables as
\begin{equation}
\left(\, \varphi(x) ,\, \chi(x) \,\right)
\equiv  (\, \tvarphi(\calA^{-1} x),\, \tchi(\calA^{-1} x) \,)
= (\,\tvarphi(\xi),\, \tchi(\xi)\,) \,,
~~~~~~\tomega \equiv \omega/\calA
\end{equation}
and using Eq.(\ref{shift}),
we can rewrite Eq.(\ref{fieldeq}) as
\begin{equation}
\left[-\frac{\partial^2}{\partial \xi^2}
- \frac{\partial \calM(\xi)}{\partial \xi} + \calM^2(\xi) \right] 
f(\xi)
= \tomega^2 \, f(\xi)\,,
~~~~~~~~
f(\xi) = \left(\, \tvarphi(\xi),\, \tchi(\xi\pm\bfK)\, \right)\,.
\end{equation}
Note that $\tvarphi(\xi)$ and $\tchi(\xi \pm\bfK)$
satisfy the same equations,
so one of the solutions can be related to
the other
by shifting the coordinate by $\bfK$,
modulo the relative phase factor.

(iv) The derivatives of the elliptic functions
are given by \footnote{See 
also the Table (16.16) in Ref. \cite{formula1}.}
\begin{equation}
\frac{\,\rmd\, \sn(\xi)\,}{\,\rmd \xi \,} 
= \cn(\xi) \, \dn(\xi)\,,
~~~
\frac{\, \rmd\, \cn(\xi)\,}{\,\rmd \xi \,} 
= - \sn(\xi) \, \dn(\xi)\,,
~~~
\frac{\, \rmd\, \dn(\xi) \,}{\,\rmd \xi \,} 
= -\lambda \, \sn(\xi) \,\cn(\xi)\,,
\end{equation}
with which one gets
\begin{equation}
\frac{\,\rmd\, \calM(\xi)\,}{\,\rmd \xi \,}
= - 2 \lambda \, \sn^2(\xi) + \lambda + \calM^2\,.
\end{equation}
Finally, we arrive at the Lame form of 
the eigenvalue equation:
\begin{equation}
\left[\, -\frac{\partial^2}{\partial \xi^2}
+ 2\lambda \, \sn^2(\xi) \, \right]
f (\xi)
= \left(\tomega^2 + \lambda\right) f (\xi) \,.
\end{equation}
The number $2$ in front of $\sn^2(\xi)$ 
is the special case of $l(l+1)$.
For given $l$,
the equation has $2l$ gaps in the energy spectra
\cite{analysis}.

\subsection{Eigenvalues and eigenfunctions}
\label{sub:ee}

To study the eigenstates,
let us first note that
the period of the potential is
$2\bfK(\lambda)$.
Therefore the eigenfunction must take
the Bloch form:
\begin{equation}
\Psi_{Q} (\xi) = v_{Q} (\xi) \, \rme^{\rmi \tQ \xi}\,,
~~~~ v_{Q} (\xi) = v_{Q} (\xi+2\bfK)\,,
\end{equation}
where $\tQ = Q/\calA$ is the (dimensionless)
quasimomentum
which is a real, continuous variable.
On the other hand, the Fourier modes
for $v_{Q} (\xi)$ can take only discrete values,
$n\pi/\bfK$, where $n$ is an integer.
Note also that the equation is the second-order
differential one, and
its kernel is real,
so that we have a pair of solutions
$(\tvarphi, \tchi)$,
and $(\tvarphi^*, \tchi^*)$.

Explicitly,
the solution of the Lame equation for $l=1$
is given by \cite{analysis,Kusnezov}
\begin{equation}
f_\alpha \left(\xi|\lambda \right) 
= \frac{\, \theta_1\left(u_{\xi+\alpha} , q |\lambda \right)\,}
{\theta_4 \left(u_{\xi} , q |\lambda \right) } \,
\rme^{ \xi Z(\alpha|\lambda) } \,,
~~~~~~
f_\alpha^* \left(\xi|\lambda \right) 
= \frac{\, \theta^*_1\left(u_{\xi+\alpha} , q |\lambda \right)\,}
{\theta_4 \left(u_{\xi} ,q |\lambda \right) } \,
\rme^{ \xi Z^*(\alpha|\lambda) } \,,
\end{equation}
where $\theta_{a}(u_{\xi}, q)$ and $Z(\alpha|\lambda)$ 
are Jacobi ellliptic theta and zeta functions
with the modulus $\lambda$,
and the variables $u_{\xi}$ and $q$ (called ``nome'') are
\begin{equation}
u_{\xi} \equiv \frac{\pi \xi}{\, 2 \bfK(\lambda) \,}
~
\rightarrow ~\left\{
\begin{matrix}
~~~~\xi~~~~~~~(\lambda \rightarrow 0)\\
~~ \frac{\pi \xi}{\,\ln \frac{16}{\lambda_1} \,
}~~~~(\lambda \rightarrow 1)
\end{matrix}
\right. \,,
~~~~~
q \equiv \rme^{-\pi \bfK'/\bfK}
~
\rightarrow ~\left\{
\begin{matrix}
~~~\frac{\lambda}{\, 16\,}~~~~~~~(\lambda \rightarrow 0)\\
~\rme^{-\frac{\pi^2}{ \ln (16/\lambda_1) } }~~(\lambda \rightarrow 1)
\end{matrix}
\right. \,,
\end{equation}
where $\bfK' = \bfK'(\lambda) = \bfK(\lambda_1)$.
Below we convert the abstract expressions into
physical notions.

(i) The parameter $\alpha$ is directly related to
 the energy spectra by the following relation:
\begin{equation}
\dn^2 (\alpha|\lambda) 
= \tomega^2  \,.
\label{alphadispersion}
\end{equation}
As we shall discuss below, there 
is restriction on the values of $\alpha$,
so $\dn^2(\alpha|\lambda)$ cannot take
arbitrary values.
Accordingly, there are forbidden regions for $\tomega$
which appear as the energy gaps in the spectra.

(ii) We can identify the Bloch periodic function
and quasimomentum as
\begin{equation}
v_{Q} (\xi) 
\equiv \frac{\, \theta_1\left(u_{\xi+\alpha}, q |\lambda \right)\,}
{\theta_4 \left(u_{\xi}, q |\lambda \right) } \, 
\exp \frac{\, -\rmi \pi \xi\,}{2\bfK} \,,
~~~~~~~
\tQ \equiv - \rmi  Z(\alpha|\lambda) + \frac{\, \pi \,}{\, 2\bfK \,} \,,
\end{equation}
which satisfy the condition $v_{Q} = v_{Q+2\bfK}$.
To understand this decomposition, we note that
the series expansions
for Jacobi ellliptic theta functions are
\footnote{Formula (16.27) in Ref. \cite{formula1}.}
\begin{align}
\theta_1(u,q) 
&= 2 q^{1/4} \sum_{n=0}^{\infty} (-1)^n q^{n(n+1)} \sin(2n+1)u\,,
~~~~~~~\theta_1(u) =\theta^*_1(u^*)
\nonumber \\
\theta_4(u,q) 
&= 1 + 2 \sum_{n=1}^{\infty} (-1)^n q^{n^2} \cos2nu \,,
\label{thetaseries}
\end{align}
from which we can verify
\begin{equation}
\theta_1 (u_{\xi} + u_{2\bfK} ) 
= \theta_1(u_{\xi} + \pi) = -\theta_1(u_{\xi}) \,,
~~~~~~~
\theta_4 (u_{\xi} + u_{2\bfK} ) = \theta_4(u_{\xi}) \,. 
\label{peritheta}
\end{equation}
The sign flipping in the first relation is the reason why
we had to include $\rme^{-\rmi \pi \xi/2\bfK}$
in $v_Q$.

\begin{figure}[tb]
\vspace{-0.5cm}
\begin{center}
  \includegraphics[scale=.30]{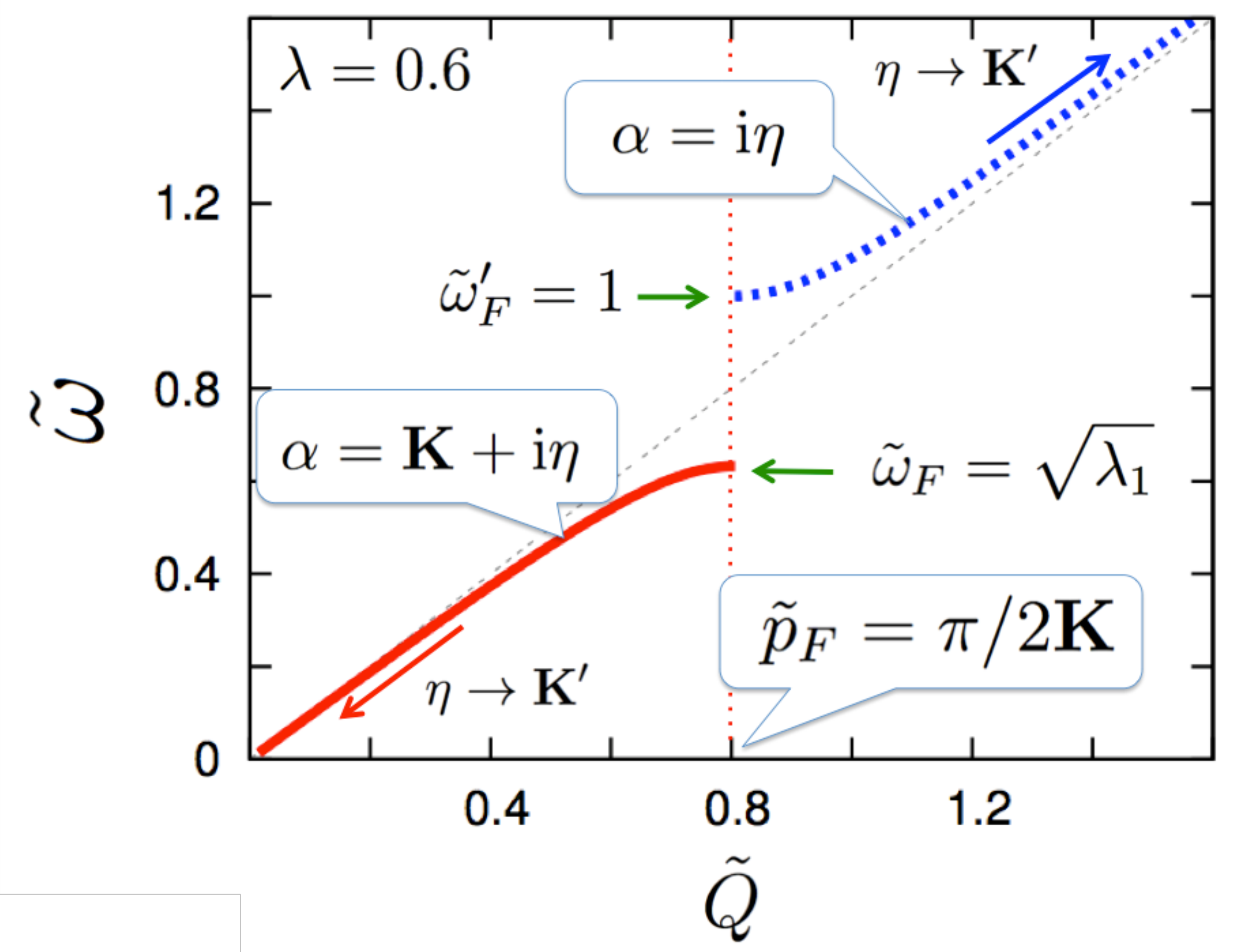} 
\end{center}
\vspace{-0.6cm}
\caption{The dispersion relation 
between the energy $\tomega$ and quasimomentum $\tQ$
(normalized by $\calA$).
$\lambda$ is chosen to be $0.6$.
The gap is opened at the quasimomentum
$\pi/2 \bfK$ ($\eta=0$),
which should be assigned as $\tp_F$.
The energy at the band edge is 
$\tomega_F = \sqrt{\lambda_1}$ ($\tomega_F' = 1)$
for the first (second) energy branch. 
The plot is symmetric with respect to 
$\tomega \rightarrow -\tomega$, and the Dirac
sea also has an energy gap of the same size.
}
\label{fig:dispersion}
\end{figure}

(iii) The quasimomentum $\tQ$ must be a real variable,
so $Z(\alpha|\lambda)$ must be purely imaginary.
This constrains the value of $\alpha$.
The series expansion of the zeta function takes the form
\footnote{Formula (17.4.38) in Ref. \cite{formula1}.}
\begin{equation}
Z(\alpha|\lambda)
= \frac{2\pi}{\,\bfK\,}
\sum_{n=1}^{\infty} \frac{ q^{n} }{\, 1-q^{2n} \, }\,
\sin \left(\frac{\, n \pi \,}{\bfK} \, \alpha \right)\,,
\label{zetaseries}
\end{equation}
which has periodicity $2\bfK$
in $\alpha$.
Thus $\alpha$ must take the form \footnote{When
$\eta$ exceeds $\bfK'$,
$q \rme^{\pi \eta/\bfK}= \rme^{\pi (-\bfK'+\eta)/\bfK} >1$, 
and power series in Eq.(\ref{zetaseries}) blow up.
Thus, $\eta \in [\, 0, \,\bfK' \,]$.}
\begin{equation}
\alpha = \rmi \eta \,,
~~~~~~\alpha = \bfK +
 \rmi \eta\,,
~~~~~~\eta \in [\, 0, \,\bfK' \,]\,.
\end{equation}
Note that at $\eta=0$ we have
$Z(0) = Z(\bfK) = 0$,
meaning that 
the quasimomenta of two branches coincide.
This is the momentum where the
energy gap appears; see Fig. \ref{fig:dispersion}.
As we will see later, to minimize
the energy of the system,
the quasimomentum at the gap
should be taken to be $p_F/\calA$, 
\begin{equation}
\tQ= \pi/2\bfK = p_F /\calA \,, ~~~~(\eta=0)
\end{equation}
so that the first positive energy branch is perfectly filled
while the second positive energy branch is empty.
This determines $\calA$ as a function of $\lambda$.

(iv) The energy branches are determined as follows.
We first examine $\alpha=\rmi \eta$ (second energy branch).
Using Jacobi's imaginary transformation
\footnote{Formula (16.20) in Ref. \cite{formula1}.},
\begin{equation}
\dn(\rmi \eta|\lambda) 
= \frac{\, \dn(\eta|\lambda_1)\,}{\, \cn(\eta|\lambda_1)\,}\,,
\label{eq:imaginary}
\end{equation}
(note that the modulus on the rhs is $\lambda_1=1-\lambda$),
we arrive at an equation for the 
second energy branch,
\begin{equation}
\tomega^2 
= \frac{\, \dn^2(\eta|\lambda_1)\,}{\, \cn^2(\eta|\lambda_1)\,}
= \lambda_1 + \frac{\, 1-\lambda_1 \,}{\, \cn^2(\eta|\lambda_1) \,}
~ \ge~ 1 ~\equiv~ \tomega_F'^{\,2}\,.
\end{equation}
Next, we examine $\alpha=\bfK +\rmi \eta$ (first energy branch).
Using  the relation for the quarter period (\ref{eq:quarter})
and then Jacobi's imaginary transformation (\ref{eq:imaginary}),
we get
\begin{equation}
\dn(\bfK+\rmi \eta|\lambda )
= \frac{\lambda_1^{1/2} }{\, \dn(\rmi \eta|\lambda) \,}
= \lambda_1^{1/2} 
\frac{\, \cn(\eta|\lambda_1)\,}{\, \dn(\eta|\lambda_1) \,} \,,
\end{equation}
and then we arrive at an equation for 
the first energy branch,
\begin{equation}
\tomega^2 
= \lambda_1 \frac{\, \cn^2(\eta|\lambda_1)\,}{\, \dn^2(\eta|\lambda_1) \,}
= 1 - \frac{\, 1- \lambda_1 \,}{\,\dn^2(\eta|\lambda_1)\,} 
~\le ~ \lambda_1  ~\equiv~ \tomega_F^2 \,.
\end{equation}
Therefore we find the energy gap
between edges of the two branches,
$\tomega_F^2 = \lambda_1$ and $\tomega_F'^{\,2}=1$.
Because the relation is given for $\tomega^2$,
we have the gaps not only near the Fermi points
but also in the Dirac sea.
The energies as functions of quasimomenta 
are plotted in Fig.\ref{fig:dispersion}.

(v) In the following calculations, we assign
eigenfunctions for $\tvarphi$ and $\tvarphi^*$ as
\begin{equation}
\tvarphi_\omega (\xi)
= \calN \frac{\, \theta_1\left(u_{\xi+\alpha} , q |\lambda \right)\,}
{\theta_4 \left(u_{\xi} , q |\lambda \right) } \,
\rme^{ \xi Z(\alpha|\lambda) } \,,
~~~~~~
\tvarphi_\omega^* (\xi)
= \calN^* \frac{\, \theta_1\left(u_{\xi+\alpha^*} , q |\lambda \right)\,}
{\theta_4 \left(u_{\xi} ,q |\lambda \right) } \,
\rme^{ -\xi Z(\alpha|\lambda) } \,,
\label{eq:phifix}
\end{equation}
where $\calN$ is the normalization factor.
The relation between these two functions
is like that between $\rme^{\rmi kx}$ and $\rme^{-\rmi kx}$
in a free fermion theory.
On the other hand, they are related to $\chi_\omega$
and $\chi_\omega^*$ through Eq.(\ref{relation}).
Actually the results in this paper
do not require the expression of the
relative phase factor.
But we give the result for completeness,
and it is given by
(for the derivation, see Appendix.\ref{sec:relaphase})
\begin{equation}
\tchi_\omega (\xi) 
= \rme^{ \rmi \Phi(\omega)} \, \tvarphi_\omega (\xi - \bfK) \,,
~~~~~~~
\rme^{ \rmi \Phi(\omega)} \equiv
{\rm sgn}(\tomega)
\, \rme^{\bfK Z(\alpha)}\,,
\end{equation}
where $\Phi$ is real and
the function $\chi(\xi)$ is proportional to 
$\varphi(\xi-\bfK)$, as stated earlier.
Note also that the phase factor changes the sign
for $\omega \rightarrow -\omega$,
as we can see from Eq.(\ref{relation}).

(vi) Finally we fix the normalization.
Since the wave function has periodicity of $2\bfK$,
the normalization condition is
\begin{equation}
1 = \frac{\, 1 \,}{\, 2\bfK \,}
\int_0^{2\bfK} \rmd \xi\,
~\left(\, |\tvarphi(\xi)|^2 + |\tchi(\xi)|^2 \,\right)
= \frac{\, 1 \,}{\, 2\bfK \,}
\int_0^{2\bfK} \rmd \xi\,
\left(\, |\tvarphi(\xi)|^2 + |\tvarphi(\xi-\bfK)|^2 \,\right) \,.
\end{equation}
We will give the explicit form of $\calN$ in 
Appendix.\ref{sec:normalization}.
Instead, here we give only the normalized expression 
for $|\tvarphi|^2$,
\begin{equation}
|\tvarphi_\omega (\xi)|^2
= \frac{1}{\, 2 \,} \left[\, 1 
- \frac{\,\dn^2 (\xi) - \bfE/\bfK \,}{\, \tomega^2 - \bfE/\bfK\,}  
\,\right]\,,
\label{normalized}
\end{equation}
where $\bfE=\bfE(\lambda)$
is the complete elliptic integral of the second kind
\footnote{Formulas (17.2), (17.3) 
and the figure (17.2) in Ref. \cite{formula1}.},
\begin{equation}
\bfE (\lambda) 
= \int^{\pi/2}_0 \!
\rmd \theta\, \sqrt{1-\lambda \sin^2 \theta \,} 
= \int_0^{\bfK} \! \rmd \xi ~ \dn^2 (\xi|\lambda)
~\rightarrow ~\left\{
\begin{matrix}
~~\pi/2~~(\lambda \rightarrow 0)\\
~~~ 1~~~~(\lambda \rightarrow 1)
\end{matrix}
\right. \,.
\label{eq:secondkind}
\end{equation}
Note that $\int_0^{2\bfK} \rmd \xi \, [\dn^2(\xi) -\bfE/\bfK]= 0$,
so the spatial average of $|\tvarphi|^2$
is saturated by the first term in 
Eq.(\ref{normalized}).

\subsection{Density of states}
\label{sub:ds}

In various computations 
we will use the density of states.
We take a derivative for the quasimomentum,
\begin{equation}
\rmd \tQ = \frac{\rmd \tQ}{\, \rmd \tomega \,}\, \rmd \tomega 
=
\frac{\rmd \tQ}{\, \rmd \alpha \,}
\frac{\rmd \alpha}{\, \rmd \tomega \,}
\, \rmd \tomega \,,
~~~~~
\calD(\omega) \equiv \left| \frac{\rmd \tQ}{\, \rmd \tomega \,} \right|\,.
\end{equation}
The $\alpha$ and $\tomega$ are related through
the relation (\ref{alphadispersion}). 
Let us first note that
\begin{equation}
\frac{\rmd}{\, \rmd \tomega \,}\, \dn(\alpha)
= \frac{\rmd}{\, \rmd \tomega \,}\, \tomega 
~~~\leftrightarrow~~~
\frac{\rmd \alpha }{\, \rmd \tomega \,}
= \frac{1}{\, \lambda \,}
\frac{ 1 }{\, \sn(\alpha)\, \cn(\alpha) \,}\,,
\end{equation}
where either $\sn(\alpha)$ or $\cn(\alpha)$
becomes purely imaginary.
Next we deal with $\rmd \tQ/\rmd \alpha$.
Taking a derivative of the dispersion relation
(see Appendix.\ref{sec:derivativezeta}),
we find
\begin{equation}
\frac{\rmd \tQ}{\, \rmd \alpha \,}
= -\rmi \frac{\, \rmd Z(\alpha)\, }{\, \rmd \alpha \,}
= -\rmi \left( \dn^2(\alpha) - \frac{\bfE}{\bfK} \right) \,.
\label{eq:zetaderivative}
\end{equation}
Assembling all these pieces,
we arrive at
\begin{equation}
\calD(\omega) 
= \mp \,
\frac{\, \tomega^2 - \bfE/\bfK \, }
{\sqrt{(\tomega^2-1) (\tomega^2- \lambda_1 )\,} \,}\,.
~~~~
\left(-~ {\rm for}~ 0 \le \, \tomega^2 \le \lambda_1\,,
~+~{\rm for}~ 1 \le \tomega^2 \,\right)
\end{equation}
Note that $\tomega^2=\bfE/\bfK$ 
is in the forbidden region
between the first and second energy branches.
In fact there is an inequality
$\lambda_1 \le \bfE/\bfK \le 1$
which can be derived
by noting that $\lambda_1 \le \dn^2x \le 1$
and $\int^{\bfK}_0 \rmd x~ \dn^2(x) = \bfE$.
Note that the density of states
is enhanced near the band edges.

\subsection{Mapping of the UV cutoff}
\label{sub:mu}

Finally, we relate the UV cutoff.
Details will be given in Appendix.\ref{sec:mapping}, 
but we will give the outline here.
First, we notice that 
$\dn(\alpha)$ for $\alpha =\rmi \eta$
approaches $+\infty$
as $\alpha \rightarrow \rmi \bfK'(\lambda)$.
Introducing an infinitesimal quantity $\epsilon$,
our energy cutoff for the inhomogenous phase,
$\omega_\Lambda$, can be expressed as
\begin{equation}
\omega_\Lambda /\calA
=  \dn\left(\, \rmi(\bfK'-\epsilon) |\lambda \, \right) \,.
\label{omegalambda}
\end{equation}
On the other hand, the number of states in the Dirac sea
is limited by $k=\Lambda$.
Using the expression for the quasimomentum, we can write
the momentum cutoff as
\begin{equation}
\Lambda /\calA
= - \rmi  Z \left(\, \rmi(\bfK'-\epsilon)|\lambda \, \right) 
+ \frac{\, \pi \,}{\, 2\bfK(\lambda) \,} \,.
\end{equation}
Expanding these equations by $\epsilon$,
we can eliminate $\epsilon$ and then relate
the momentum cutoff to the energy cutoff,
\begin{equation}
\omega^2_\Lambda 
= \Lambda^2 
+ \calA^2\, \left[\, 
- \lambda + 2 \left(\, 1 - \frac{\bfE}{\, \bfK \,} \,\right) 
\, \right]
+ O(\Lambda^{-2})\,.
\label{momenergycutoff}
\end{equation}
The $O(1)$ terms must be kept 
during the following calculations.

\section{Expectation values}
\label{sec:condensates}

Now we have all the ingredients
to compute various quantities.
We first write down expressions for the
fermion number density and energy density,
and then determine the variational parameters
$\calA$ and $\lambda$ as functions of
average density or $p_F$.
After that we compute the spatial modulations
of various density operators:
fermion number, energy, and scalar and pseudoscalar density.
At the end of this section, in Sec. \ref{limits},
we examine the high and low density limits
of various quantities to get qualitative insights.

Using the bases found in the previous section,
the fermion field operator 
can be written in terms of the creation and annihilation operators,
\begin{equation}
\hat{\psi}_{c} (x) 
= \int \frac{\, \rmd \omega \,}{2\pi}\, 
\sqrt{\, \calD(\omega) \,}\,
\sum_{j=1,2} u^j_\omega (\xi)
\left[\, \theta(\omega)\, 
\rme^{-\rmi \omega t}\, \hat{a}_c(\omega,j)  
+ \theta(-\omega) \, \rme^{\rmi \omega t}\, 
\hat{b}_c^\dag (\omega,j) \, \right]\,,
\label{eq:fieldop}
\end{equation}
where $\hat{a}$ and $\hat{b}$ are annihilation operators
for particles and antiparticles,
and $c$ is used for color indices.
$\calD(\omega)$ is the density of states,
and we took the normalization of
the creation and annihilation operators
to satisfy 
$\{a_c(\omega,j), a_{c'}^\dag(\omega',j')\}
=2\pi \delta(\omega-\omega') \delta_{jj'} \delta_{cc'}$.
[This normalization requires
$\sqrt{\calD}$ in Eq. (\ref{eq:fieldop}).]
The wave functions giving energy $\omega$ are
\begin{equation}
u_\omega^1 (\xi) = 
\left[\begin{matrix}
\, \tvarphi_\omega (\xi) \, \\
\, \tchi_\omega (\xi) \,
\end{matrix} \right]\,,
~~~~~
u_\omega^2 (\xi) = 
\left[\begin{matrix}
\, \tvarphi^*_\omega (\xi) \, \\
\, \tchi^*_\omega (\xi) \,
\end{matrix} \right]\,,
~~~~~
\tvarphi_\omega = \tvarphi_{-\omega}\,,
~~~~~
\tchi_\omega = - \tchi_{-\omega}\,,
\label{wf}
\end{equation}
where for the convention 
$\tvarphi_\omega = \tvarphi_{-\omega}$,
the relative sign in $\tchi_\omega$
for the positive and negative 
energy accompanies $(-1)$.

\subsection{Fermion number:
Determination of $\calA$}
\label{fermionnum}

The fermion number density is given by
\begin{equation}
\left\la \psi^{\dag} \psi \right\ra
= \int\! \frac{\, \rmd \omega \,}{2\pi} 
\, \calD(\omega) \sum_{j=1,2} |u^j_\omega (\xi)|^2
\left\la \theta(\omega)\, 
\hat{a}^\dag(\omega,j)  \hat{a}(\omega,j)  
+ \theta(-\omega)
\left[\, 1 -\hat{b}^\dag(\omega,j)  \hat{b}(\omega,j) \right] 
\right\ra \,,
\end{equation}
where the sum over color indices is implicit.
Considering the fermion number constraint 
and the fact that the Dirac sea does not contain
any antiparticles,
we may require
\begin{equation}
\left\la \hat{a}^\dag(\omega,j)  \hat{a}(\omega,j)  \right\ra
= N\, \theta(\epsilon_F -\omega) \,,
~~~~~
\theta(-\omega)
\left\la \hat{b}^\dag(\omega,j)  \hat{b}(\omega,j)  \right\ra 
=0\,,
\end{equation}
where $\epsilon_F$ is the Fermi energy which will
be fixed below.
We arrive at
\begin{align}
\left\la \psi^{\dag} \psi(x) \right\ra
&= 2N \left(\int_0^{\epsilon_F} + \int_{-\omega_\Lambda}^0\right)
 \! \frac{\, \rmd \omega \,}{2\pi} \,\calD(\omega) 
\left(\, |\tvarphi_\omega (\xi)|^2 +  |\tchi_\omega (\xi)|^2 \,
\right) \nonumber \\
&=N
\left(\int_0^{\epsilon_F} +\int_{-\omega_\Lambda}^0\right)
 \! \frac{\, \rmd \omega \,}{2\pi} \, \calD(\omega)
\left[\, 2 - 
\frac{\, \calF(\xi) \,}
{\omega^2/\calA^2 - \bfE/\bfK } \,\right]
\,.
\end{align}
where we have defined
\begin{equation}
\calF(\xi) \equiv  \dn^2(\xi) + \dn^2(\xi-\bfK) - 2 \bfE/\bfK \,,
\end{equation}
whose spatial average over the period $2\bfK$ is zero.
Here we took into account the particles
which fill the Dirac sea.
The average part in the Dirac sea
will be eliminated by the vacuum subtraction,
while the spatial modulation is not
and requires some care.

(i) Average density: 
We first compute the constant part.
In order to minimize the energy,
the particle should fill the first valence band,
leaving the upper energy branch empty.
In Appendix.\ref{sec:location},
we will show that locating
the Fermi surface at the gapped points
indeed reduces the energy density.
Therefore, in the following, we set
$\epsilon_F = \omega_F = \sqrt{\lambda_1} \calA$.
Next, notice that we impose the UV cutoff on momenta,
so the size of phase space in the Dirac sea is kept fixed.
Therefore, the Dirac sea contribution to the femion number density
is common for all phases.
Thus, subtracting the Dirac sea contribution,
we demand that
\begin{equation}
\frac{\, \left\la \psi^{\dag} \psi(x) \right
\ra_{ {\rm ave.}  }^{ R } \,} {N}
= \int_0^{\omega_F=\sqrt{\lambda_1} \calA } 
\frac{\, \rmd \omega \,}{\pi}
\frac{\,  \calA^2 \bfE/\bfK -\omega^2\,}
{\sqrt{(\omega^2-\calA^2) (\omega^2- \lambda_1 \calA^2)\,} \,}
= \frac{\, p_F \,}{\pi} \,.
\end{equation}
Taking the variable $\omega = \sqrt{\lambda_1} \calA t$,
the integral can be expressed as
\begin{equation}
\frac{\calA}{\pi} \int_0^{1} \rmd t
\left[ \sqrt{\frac{\,  1-\lambda_1 t^2\,}
{\, 1-t^2\,} \,}
- 
\frac{\,1-\bfE/\bfK \,}
{\sqrt{(1-t^2)(1-\lambda_1 t^2)\,} } \right]
=\frac{\calA}{\, \pi \bfK \,}
\left[\, \bfE' \bfK + \bfE \bfK' - \bfK \bfK' \,\right]
= \frac{\, \calA \,}{\, 2\bfK \,} \,,
\end{equation}
where in the first step we used
the integral expression for 
$\bfE'=\bfE(\lambda_1) $ and $\bfK'=\bfK(\lambda_1)$,
and in the last step
we have used Legendre's relation.
Now $\calA$ is fixed to
\begin{equation}
\calA = \frac{\, 2\bfK\,}{\pi}\, p_F
~\rightarrow~ \left\{ \begin{matrix}
~~~~~~~p_F~~~~~~~~ (\lambda \rightarrow 0)\\
~~~\frac{\, p_F \,}{\pi} \ln \frac{16}{\, \lambda_1 \,}
~~~~~(\lambda \rightarrow 1)
\end{matrix} \right.
\label{solA}
\end{equation}
Now the only remaining variational parameter
is $\lambda$.

(ii) The spatially modulating part:
Next, we treat the spatially modulating part
(whose spatial average is zero).
It is given by (see Fig. \ref{fig:dispersion} for a reminder)
\begin{equation}
- N \frac{\, \calA^2 \,}{\, 2\pi \,}
\left(\int_0^{\omega_F} 
+ \int_{-\omega_F}^0
+ \int_{-\omega_\Lambda}^{-\omega'_F} \right)
 \! \rmd \omega \, 
\frac{ {\rm sgn} 
\left(\omega^2 - \calA^2 \bfE/\bfK \right) }
{\sqrt{(\omega^2-\calA^2) (\omega^2- \lambda_1 \calA^2)\,} }
\, \calF(\xi) \,.
\end{equation}
The integral part from the first energy
branches in the Fermi (Dirac) sea 
gives
\begin{equation}
-\int_0^{\omega_F} \!
\frac{\, \rmd \omega \,}
{\, \sqrt{(\omega^2-\calA^2) (\omega^2-\lambda_1 \calA^2)\,} \,}
= -\frac{1}{\, \calA \,}
\int_0^{1} 
\frac{\rmd t}{\, \sqrt{(1-t^2)(1-\lambda_1 t^2)\,} \,} 
= -\frac{\, \bfK' (\lambda) \,}{\calA}\,,
\label{eq:firstnum}
\end{equation}
where we have changed the variable
as $\omega = \sqrt{\lambda_1} \calA t$.
On the other hand, the second energy branch in the Dirac sea gives
a contribution with the same size but opposite sign
($\omega_F'=\calA$),
\begin{equation}
\int_{-\omega_\Lambda}^{-\calA} 
\frac{\,\rmd \omega \,}
{\, \sqrt{(\omega^2-\calA^2) (\omega^2-\lambda_1 \calA^2)\,} \,}
= \frac{\, \bfK' (\lambda) \,}{\calA}\,.
\end{equation}
This can be checked by noting that 
the change of the variable 
$\omega \rightarrow 1/\omega'$
converts the integral into 
the same form as that in Eq.(\ref{eq:firstnum}).
Note that the Dirac sea contributions from the first 
and the second energy branches cancel out,
leaving only the net contribution from the Fermi sea.

Assembling the spatial average and modulating parts,
the fermion number density is given by
\begin{equation}
\left\la \psi^{\dag} \psi(x) \right\ra
= N \, \frac{\, p_F \,}{\pi} 
\left(\, 1 + \frac{\, \bfK \bfK' \,}{\, \pi \,}\, \calF(\xi) \right)\,,
~~~~~(\xi=\calA x)
\end{equation}
where we have used 
$\calA = 2p_F \bfK/\pi$ in Eq.(\ref{solA}).
The behavior at $\lambda =0.9$ is plotted in
Fig. \ref{fig:spirals}.

\subsection{Energy density:
Determination of $\lambda$}
\label{fermionnum}

Next we compute the energy density.
The single-particle energy contribution
gives
\begin{align}
\calE_1(x) 
& \equiv 2N
\left(\int_0^{\omega_F} +\int_{-\omega_\Lambda}^0\right)
 \! \frac{\, \rmd \omega \,}{2\pi} \, \calD(\omega) \, \omega \,
\left(\, |\tvarphi_\omega (\xi)|^2 +  |\tchi_\omega (\xi)|^2 \, \right)
\nonumber \\
&= - N
\int_{\omega_F'}^{\omega_\Lambda} 
 \! \frac{\, \rmd \omega \,}{2\pi} \, \calD(\omega) \, \omega
\left[\, 2 - 
\frac{\, \calF(\xi) \,}
{\omega^2/\calA^2 - \bfE/\bfK } \,\right]
\,,
\end{align}
where the integrand is an odd function of $\omega$,
so that Fermi and Dirac sea 
contributions from the first energy branches cancel,
leaving the contribution 
from the second energy branch in the Dirac sea.
(We have changed the variable as $\omega\rightarrow -\omega$.)
Writing the spatial average and spatial fluctuation parts
as $\bar{\calE}_1$ and $\Delta \calE$,
straightforward calculations lead to
\begin{align}
\frac{\, \bar{\calE}^R_1 \,}{N}
& = 
- \frac{\, \calA^2 \,}{\, 4\pi \,}
\left[\, 
\left( 2-\lambda  - 2\, \frac{\, \bfE \,}{\bfK} \right) 
\ln \frac{\, 4\Lambda^2 \,}{\, \lambda \calA^2 \,} 
+
\left( 2-\lambda  - 4\, \frac{\, \bfE \,}{\bfK} \right) 
\, \right]\,,
\nonumber \\
\frac{\, \Delta \calE_1 \,}{N}
&
= \frac{\, \calA^2 \,}{\, 4\pi \,} \, \calF(\xi)
\ln \frac{\, 4\Lambda^2 \,}{\, \lambda \calA^2 \,} 
\,,
\end{align}
where we define the regularized energy
$\bar{\calE}^R_1 \equiv \bar{\calE}_1 - \calE_{ {\rm vac} }$
where $\calE_{ {\rm vac} } = -N\Lambda^2/2\pi$,
and drop the $O(1/\Lambda)$ terms.

Next we consider the contribution
from the condensation terms.
We first note that
\begin{equation}
\frac{\, M^2 (x) \,}{\calA^2} 
= \lambda^2 \frac{\, \sn^2(\xi)\, \cn^2(\xi) \,}{\dn^2(\xi)}
=  
\left(2-\lambda - 2\, \frac{\bfE}{\, \bfK \,} \right)
- \calF(\xi)
\,,
\end{equation}
where the first bracket gives the spatial average,
as we can see from the second bracket
which is vanishing after averaging over the period $2\bfK$.
These terms have coefficient $1/G$
whose renormalized value is 
determined through the renormalization condition,
\begin{equation}
\frac{1}{\, G(\Lambda) \,}
= \frac{1}{\, 2\pi \,} \ln \frac{4\Lambda^2}{\, M_0^2 \,} \,,
\end{equation}
where $M_0$ is the effective mass in vacuum.
With this expression, we have 
the average and fluctuation parts of the condensation energy
($\calE_2/N \equiv M^2(x)/2G$)
\begin{equation}
\frac{\, \bar{\calE}_2 \,}{N}
 = \frac{\, \calA^2 \,}{\, 4\pi \,}
\left( 2-\lambda  - 2\, \frac{\, \bfE \,}{\bfK} \right) 
\ln \frac{\, 4\Lambda^2 \,}{\, M_0^2 \,} \,,
~~~~
\frac{\, \Delta \calE_2 \,}{N}
= - \frac{\, \calA^2 \,}{\, 4\pi \,} \, \calF(\xi)
\ln \frac{\, 4\Lambda^2 \,}{\, M_0^2 \,} 
\,.
\end{equation}
After combining $\calE_1$ and $\calE_2$,
we can erase $\Lambda$ in the logarithms,
and the energy depends on $\Lambda$
only through the renormalized paramemeter.
Now we can write down 
the average and fluctuating parts of total energy.
The average part is
\begin{equation}
\frac{\, \left( \bar{\calE}_1 + \bar{\calE}_2 \right)_R\,}{N}
 = - \frac{\, \calA^2 \,}{\, 4\pi \,}
\left[\,  \left( 2-\lambda  - 2\, \frac{\, \bfE \,}{\bfK} \right) 
\ln \frac{\, M_0^2 \,}{\, \lambda \calA^2 \,} 
+ \left( 2-\lambda  - 4\, \frac{\, \bfE \,}{\bfK} \right) 
\, \right]
\,,
\end{equation}
where $\calA=2p_F \bfK/\pi$ due to 
the fermion number constraints; see Eq. (\ref{solA}).
We have to choose the value of $\lambda$
so as to minimize the total average energy density.
Using a relation
\begin{equation}
\partial_\lambda \bfE
=\partial_\lambda \int_0^1 \rmd t 
\sqrt{\frac{1-\lambda t^2}{1-t^2} }
= \frac{\, \bfE - \bfK \,}{2\lambda}\,,
\end{equation}
we can show that only terms with the logarithmic
coefficient survive:
\begin{equation}
\partial_\lambda \left(\bar{\calE}_1 + \bar{\calE}_2 \right) =0
~~~\rightarrow ~~~
0=\ln \frac{\, M_0^2 \,}{\, \lambda \calA^2 \,} \times 
\partial_\lambda \left[\calA^2 \left( 2-\lambda  - 2\, \frac{\, \bfE \,}{\bfK}
\right) \right]\,.
\label{minimization}
\end{equation}
Thus we get a transcendental equation
from the vanishing logarithmic term,
\begin{equation}
M_0 =\sqrt{\lambda} \calA(\lambda) 
= \frac{\, 2 p_F \,}{\pi} 
\times \sqrt{\lambda} \, \bfK(\lambda)\,,
\label{transcendal}
\end{equation}
which determines the optimal $\lambda$
as a function of $p_F/M_0$.
Note that the optimized $\lambda$
makes the spatial modulating part of the energy density
vanishing,
\begin{equation}
\frac{ \Delta \left( \calE_1 +\calE_2 \right)}{N}
= \frac{\, \calA^2 \,}{\, 4\pi \,} \, \calF(\xi)
\ln \frac{\, M_0^2 \,}{\, \lambda \calA^2 \,} 
~\rightarrow~ 0\,.
\end{equation}
meaning that 
the energy density is uniform everywhere.

\subsection{Scalar density: 
The self-consistency condition}
\label{fermionnum}

The scalar density can be expressed as
\begin{equation}
\left\la \bar{\psi} \psi(x) \right\ra
= - 2N \left(\int_0^{\omega_F} 
+ \int_{-\omega_F}^0 + \int_{-\omega_\Lambda}^{-\omega_F'}
\right) 
 \! \frac{\, \rmd \omega \,}{2\pi} \,\calD(\omega) 
\left[\, \tvarphi^*_\omega \tchi_\omega (\xi) 
+  \tchi^*_\omega \tvarphi_\omega (\xi) \,
\right] \,.
\label{scalarintegral}
\end{equation}
Note that in contrast to the fermion number density,
the integrand is an odd function of $\omega$
because $\tvarphi_\omega = \tvarphi_{-\omega}$
and $\tchi_\omega = - \tchi_{-\omega}$.
As a consequence, 
the contributions from the first energy branches
in the Fermi and Dirac sea cancel,
and only the third integral in (\ref{scalarintegral})
gives the net contribution.
Therefore, in the GN model,
the net contribution to the scalar density
is dominated by the Dirac sea contribution.
We will discuss this point in more detail 
in Sec.\ref{sec:discussions}.

Changing the variable to keep the integration
domain in positive values,
we have
\begin{equation}
\left\la \bar{\psi} \psi(x) \right\ra
= 2N 
\int^{\omega_\Lambda}_{\omega'_F} 
 \! \frac{\, \rmd \omega \,}{2\pi} \,\calD(\omega) 
\left[\, \tvarphi^*_\omega \tchi_\omega (\xi) 
  +  \tchi^*_\omega \tvarphi_\omega (\xi) \,
\right] \,.
\end{equation}
We can express $\tchi$ in terms of $\tvarphi$,
and we arrive at
\begin{equation}
\tvarphi^*_\omega \tchi_\omega 
+ \tvarphi_\omega \tchi^*_\omega
= -\frac{\, \calA\,}{\omega} \big[\,
\tvarphi_\omega^*
\left( \partial_\xi  + \calM \right) \tvarphi_\omega
+ \tvarphi_\omega 
\left( \partial_\xi  + \calM \right) \tvarphi^*_\omega
\, \big]
= -\frac{\, \calA\,}{\omega} 
\big(\, \partial_\xi  + 2\calM \, \big)\, 
 |\tvarphi_\omega|^2 \,.
\end{equation}
Using Eq.(\ref{normalized}),
straightforward calculations lead to
\begin{equation}
\tvarphi^*_\omega \tchi_\omega + \tvarphi_\omega \tchi^*_\omega
= - \frac{\omega/\calA}{\, \omega^2/\calA^2 -\bfE/\bfK \,}
\,\calM(\xi) \,.
\end{equation}
Note that the scalar density at any energy level
is proportional to $\calM$.
Finally, we sum over all the levels
for the second energy branch in the Dirac sea,
\begin{equation}
\left\la \bar{\psi} \psi (x) \right\ra
 = - 2N \calA \calM(\xi) 
\int^{\omega_\Lambda}_{\omega'_F} 
 \! \frac{\, \rmd \omega \,}{2\pi} \,
\frac{\,\omega  \,}
{\, \sqrt{(\omega^2-\calA^2) (\omega^2-\lambda_1 \calA^2)\,} \,}\,,
\end{equation}
which yields
\begin{equation}
\left\la \bar{\psi} \psi (x) \right\ra
= - N\, \frac{\, \calA \calM(\xi) \,}{\, 2\pi \,}
\ln \frac{4\Lambda^2}{\, \lambda \calA^2 \,}
= - N\, \frac{\, M (x) \,}{\, G \,} 
\,.
\end{equation}
Here we have used the relation determined
by energy minimization,
$\sqrt{\lambda} \calA = M_0 
= 2\Lambda\, \rme^{- \frac{\pi}{G} }$.
The final expression proves
the self-consistent condition.
The behavior of the scalar density 
at $\lambda =0.9$ is plotted in
Fig. \ref{fig:spirals}.

\subsection{Pseudoscalar density}

Next we will investigate the pseudoscalar density.
At energy $\omega$,
we have ($\gamma_5=-\gamma^5$)
\begin{equation}
\bar{\psi}_\omega \rmi \gamma_5 \psi_\omega
= 2 N\left(\, |\tvarphi_\omega|^2 - | \tchi_\omega|^2 \,\right)\,.
\end{equation}
Note that the integrand is an even function of $\omega$
in contrast to the scalar density case.
We did similar calculations for the 
spatially modulating part of the fermion number density,
and found that the Dirac sea contributions
in the first and second energy branches cancel 
out by themselves.
The situation is similar here.
The net contribution to pseudoscalar density comes only 
from the Fermi sea,
\begin{equation}
\frac{\, \left\la \bar{\psi} \rmi \gamma_5 \psi(x) \right\ra \,}{N}
=  \frac{\, \calA^2 \,}{2\pi}  
\int_0^{\omega_F} 
 \! \rmd \omega \, 
\frac{\, \dn^2(\xi) - \dn^2(\xi-\bfK)  \,}
{\, \sqrt{(\omega^2-\calA^2) (\omega^2-\lambda_1 \calA^2)\,} \,} 
=  \frac{\, \bfK' \calA \,}{\, 2\pi \,} 
\left[\, \dn^2(\xi) - \dn^2(\xi-\bfK)  \,\right]
\,,
\end{equation}
where we have used the spectral weights which we have computed 
for the fermion number
[see Eq. (\ref{eq:firstnum})].
Now we have verified that
the pseudoscalar condensate exists in the GN model
at finite density, as stated in Introduction.
While its spatial average is zero,
it is locally nonzero in space.

Actually it is more instructive
to express the pseudoscalar density 
in another way.
Using the Dirac equation,
we can derive a relation,
\begin{equation}
\frac{\, \bar{\psi}_\omega \rmi \gamma_5 \psi_\omega \,}{N}
= 2\left(\, |\tvarphi_\omega|^2 - | \tchi_\omega|^2 \,\right)
= \frac{1}{\, \tomega \,}\,
\partial_\xi 
\left(\tvarphi_\omega \tchi_\omega^* + \tvarphi^*_\omega \tchi_\omega
\right) 
= \frac{1}{\, 2\tomega \,}\,
\frac{\, \partial_\xi \left( \bar{\psi}_\omega \psi_\omega \right)\,}{N}
\,.
\end{equation}
Therefore at given $\omega$,
the pseudoscalar density is proportional to
the spatial gradient of the scalar density.
After integrating over $\omega$ with the spectral weight,  
we find
\begin{equation}
\left\la \bar{\psi} \rmi \gamma_5 \psi (x) \right\ra
=  N \, \frac{\, \bfK' \calA \,}{\, 2\pi \,}\, \partial_\xi \calM (\xi) \,.
\end{equation}
This expression clarifies that
the {\it inhomogeneity} 
of the chiral scalar condensate
drives the formation of the
pseudoscalar condensate.
The typical behavior is shown in Fig. \ref{fig:spirals}.

\subsection{High and low density limits}
\label{limits}

\begin{figure}[tb]
\vspace{0.0cm}
\begin{center}
\scalebox{0.6}[0.6] {
  \includegraphics[scale=.40]{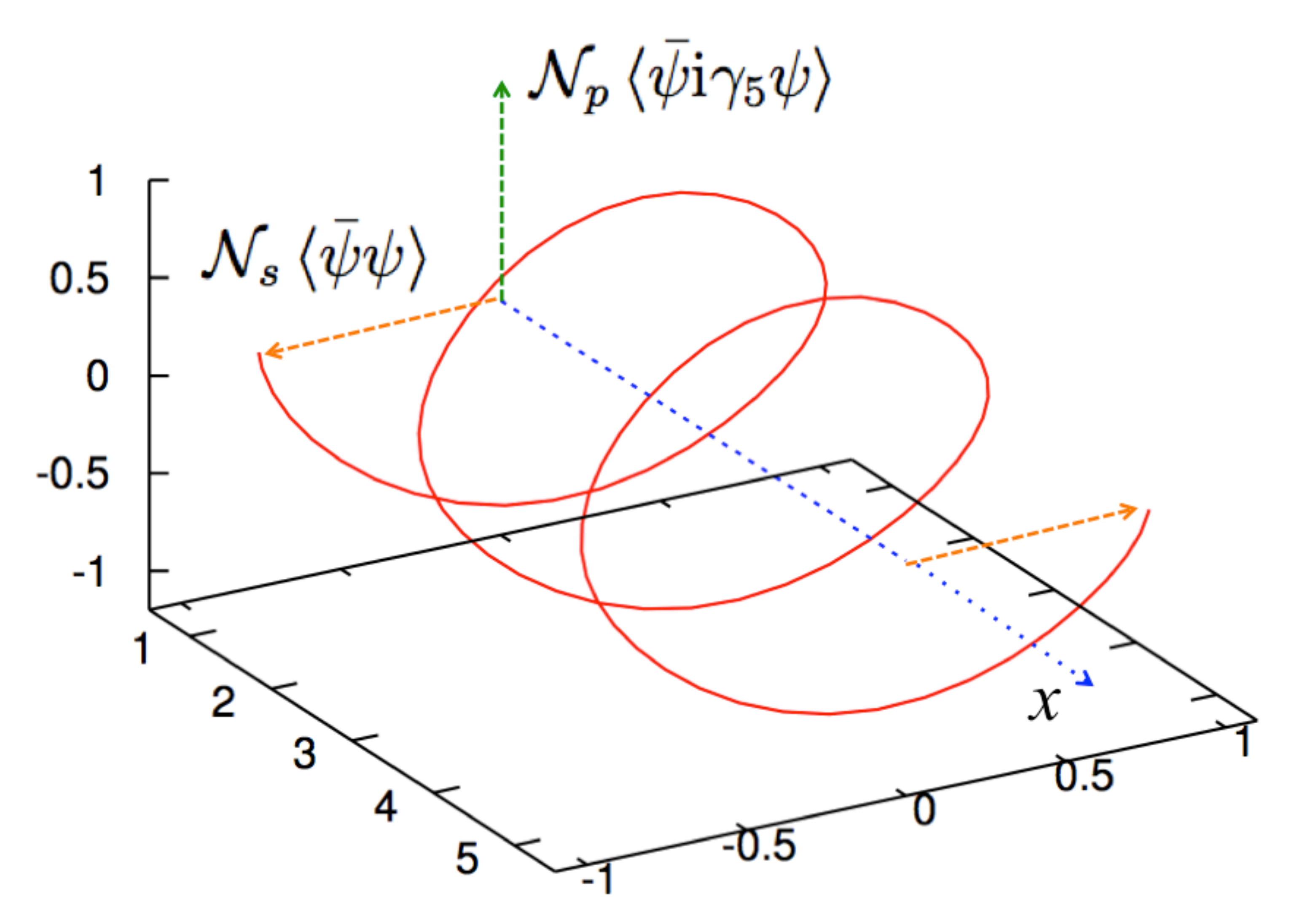} 
} \hspace{0.2cm}
\scalebox{0.6}[0.6] {
  \includegraphics[scale=.40]{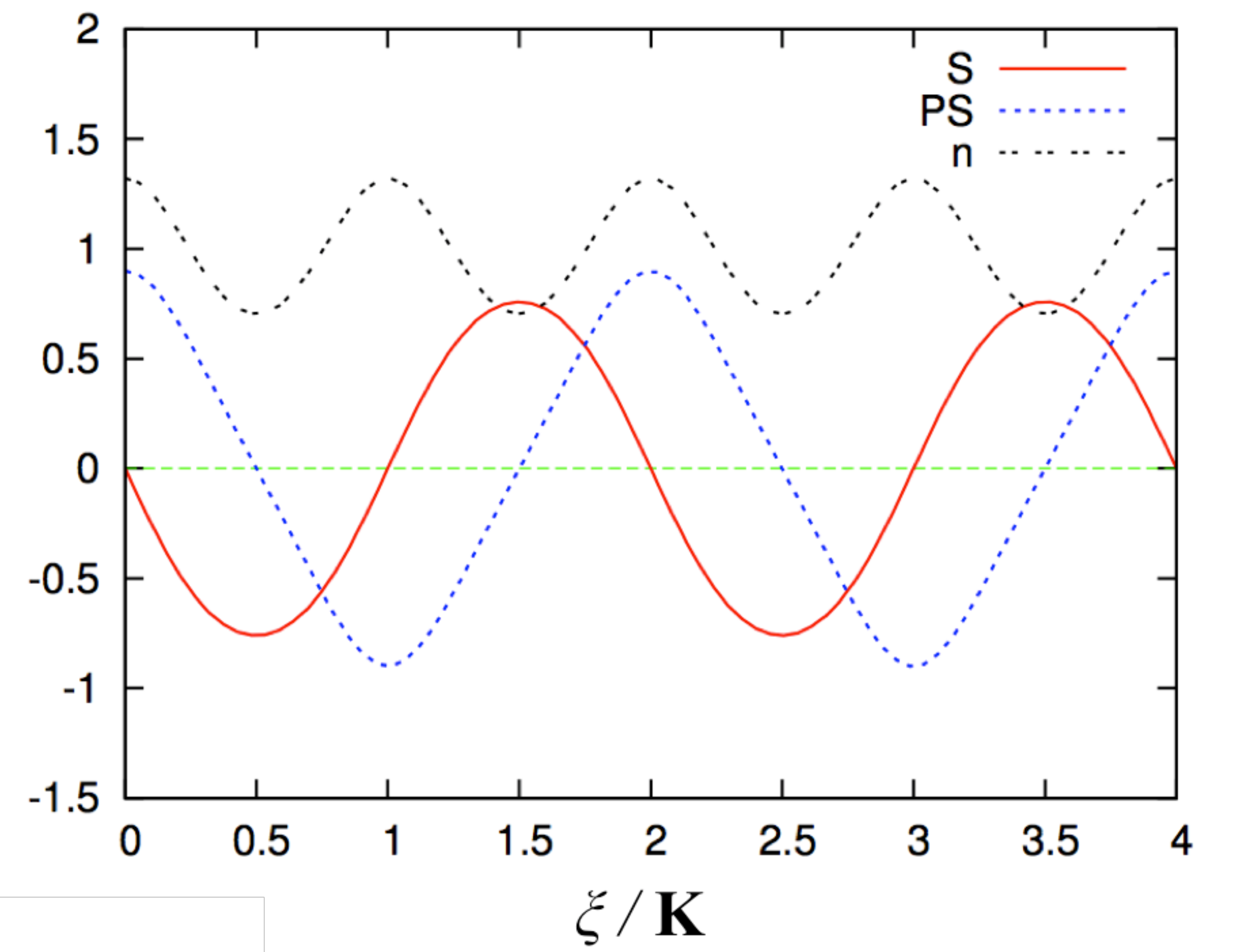} }
\end{center}
\vspace{-0.4cm}
\caption{Left: The ``elliptic'' chiral spirals at $\lambda=0.9$.
We plot the amplitude-free parts of condensates,
$\calN_s \la \bar{\psi} \psi \ra
\equiv \la \bar{\psi} \psi \ra \times G/N\calA=-\calM$
and $\calN_p \la \bar{\psi} \rmi \gamma_5 \psi \ra
\equiv \la \bar{\psi} \rmi \gamma_5 \psi \ra \times 2\pi /N \bfK'\calA
=\partial_\xi \calM$.
Right: The plots at $\lambda=0.9$ for 
the (normalized) scalar (S), pseudoscalar (PS),
and fermion number density (n).
For the fermion number density,
we divide by $Np_F/\pi$.
The fermion number is stuck at the location of the 
domain wall where the scalar density passes zero.
}
\label{fig:spirals}
\vspace{0.1cm}
\end{figure}

We consider the high and low density limits
and examine qualitative aspects of the quantities
which we have computed so far.
To begin with,
we first express $\lambda$ in terms of $p_F/M_0$.
The transcendental equation (\ref{transcendal})
in the $\lambda\rightarrow 0 $ and $\lambda \rightarrow 1$ 
cases becomes
\begin{equation}
\frac{\, \pi M_0 \,}{\, 2p_F \,}
= \sqrt{\lambda}\, \bfK(\lambda)
~\rightarrow~
\left\{
\begin{matrix}
~~~
\frac{\, \sqrt{\lambda}\,\pi}{2} 
\left[\, 1+\frac{\lambda}{4} + O(\lambda^2) \,\right]
~~~~~~~&(\lambda \rightarrow 0)\\
~~~ \frac{1}{\,2\,} \ln \frac{\, 16\,}{\, \lambda_1 \,}
+O(\lambda_1 \ln \lambda_1)
~~~~~~&(\lambda \rightarrow 1)
\end{matrix}
\right. \,,
\end{equation}
from which we get
\begin{equation}
\lambda 
=
\left\{
\begin{matrix}
~~~
\left( \frac{\, M_0\,}{\, p_F \,} \right)^2
- \frac{\,1\,}{\, 2 \,} \left( \frac{\, M_0\,}{\, p_F \,} \right)^4
+O(M_0^6/p_F^6)
~~~~~~~&( M_0/p_F \ll 1)\\
~~~~ 1 - 16 \, \rme^{ - \pi M_0/p_F  }
+ O\left(\rme^{ - 2\pi M_0 / p_F  } \right)
~~~~~~~&( 1 \ll M_0/p_F)
\end{matrix}
\right. \,.
\end{equation}
Next, we look at the parameter $\calA$
which appears in place of the coordinate, $\xi =\calA x$.
Its asymptotic behavior is given by
\begin{equation}
\calA
= \frac{\, 2\bfK(\lambda) \,}{\pi} p_F
=
\left\{
\begin{matrix}
~~~
p_F \left[\, 1+\frac{1}{\, 4\,} \left(\frac{\, M_0\,}{\, p_F \,} \right)^2 
+O(M_0^4/p_F^4) \, \right]
~~~~~~~&( M_0/p_F \ll 1)\\
~~ M_0
\, + \,O\left( p_F\, \rme^{ - \pi M_0/p_F } \right)
~~~~~~~~~&( 1 \ll M_0/p_F)
\end{matrix}
\right. \,.
\end{equation}
Now we shall consider the
physical quantities of particular interest.

(i) The asymptotic behavior of the energy gap is 
\begin{equation}
\Delta_g
\equiv \omega_F' - \omega_F 
= \left(1-\sqrt{\lambda_1} \right) \calA
= \left\{
\begin{matrix}
~~
M_0 \times \frac{\, M_0 \,}{\, 2p_F \,} + \cdots
~~~~~& ( M_0/p_F \ll 1)\\
~~ M_0 + \cdots
~~~~& ( 1 \ll M_0/p_F)
\end{matrix}
\right. \,.
\end{equation}
In particular,
at high density the gap is proportional to $1/p_F$
and tends to close rather quickly.
It is important to notice that
this quick decreasing behavior is not generic
in other two-dimensional models.
For instance, in the NJL$_2$ model 
the gap stays at the vacuum value, $\sim M_0$.
We will discuss this issue more 
in Sec.\ref{sec:discussions}.

(ii) The low density behaviors
of the chiral scalar and pseudoscalar condensates
are
\begin{equation}
\frac{\, \la \bar{\psi} \psi \ra \,}{N}
\simeq -\, \frac{\,M_0\,}{G} \tanh(M_0x) \,,
~~~~~
\frac{\, \la \bar{\psi} \rmi \gamma_5 \psi \ra \,}{N}
\simeq \, \frac{\,M_0\,}{4} \frac{1}{\, \cosh^2 (M_0x) \,}\,.
~~~( 1 \ll M_0/p_F)
\end{equation}
This is the solution for widely separated kinks.
When the scalar density becomes zero,
the pseudoscalar density is maximized.
On the other hand,
the scalar density is maximized when
the pseudoscalar density is zero.
Therefore, the combination of the scalar and pseudoscalar
density forms the chiral spirals,
as shown in Fig.\ref{fig:spirals}.

(iii) The high density behaviors
of the chiral scalar and pseudoscalar condensates
are
\begin{equation}
\frac{\, \la \bar{\psi} \psi \ra \,}{N}
\simeq -\,\frac{\,\Delta_g\,}{\, G_\Lambda \,} \sin(2p_F x) \,,
~~~~~
\frac{\, \la \bar{\psi} \rmi \gamma_5 \psi \ra \,}{N}
\simeq \frac{\,\Delta_g\,}{\, G_{2p_F} \,} \cos(2p_F x) \,,
~~~( M_0/p_F \ll 1)
\end{equation}
where $G_{2p_F}$ is defined
by substituting $2p_F$ in place of
$\Lambda$ in the coupling constant $G(\Lambda)$.
This disparity of the effective coupling constants
reflects the fact that the scalar and pseudoscalar
density acquire contributions from different domains.
We can construct an approximate invariant,
\begin{equation}
\left( G_\Lambda \left\la \bar{\psi} \psi \right\ra \right)^2
+ \left( G_{2p_F} \left\la \bar{\psi} \rmi \gamma_5 \psi \right\ra \right)^2
\simeq N^2 \Delta_g^2 \,.
\end{equation}
As $p_F$ becomes larger,
the expression approaches the chiral spirals with equal amplitudes
for the scalar and pseudoscalar density.

\section{Discussions}
\label{sec:discussions}

\begin{figure}[tb]
\vspace{0.0cm}
\begin{center}
\scalebox{0.6}[0.6] {
  \includegraphics[scale=.40]{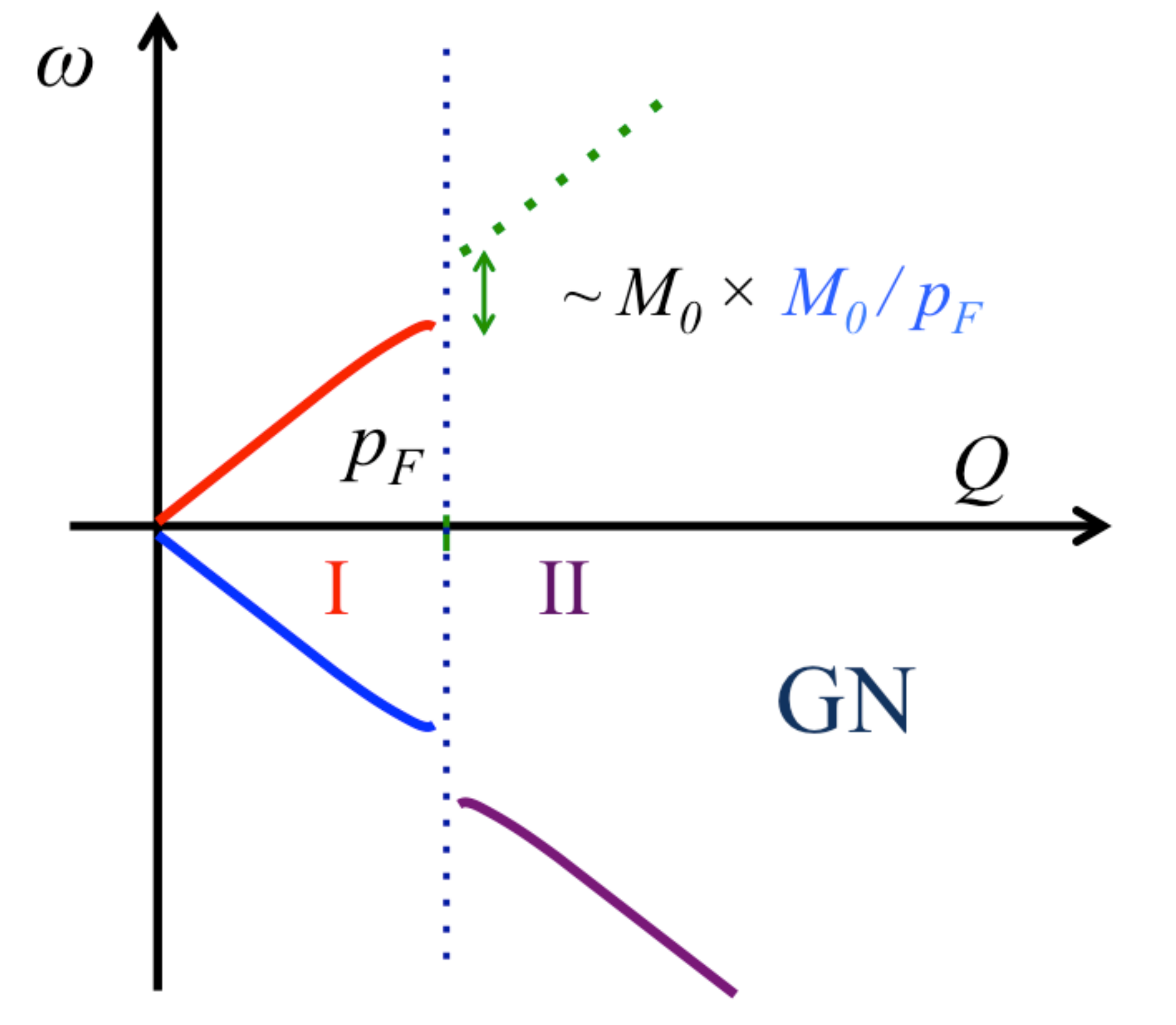} 
} \hspace{0.2cm}
\scalebox{0.6}[0.6] {
  \includegraphics[scale=.40]{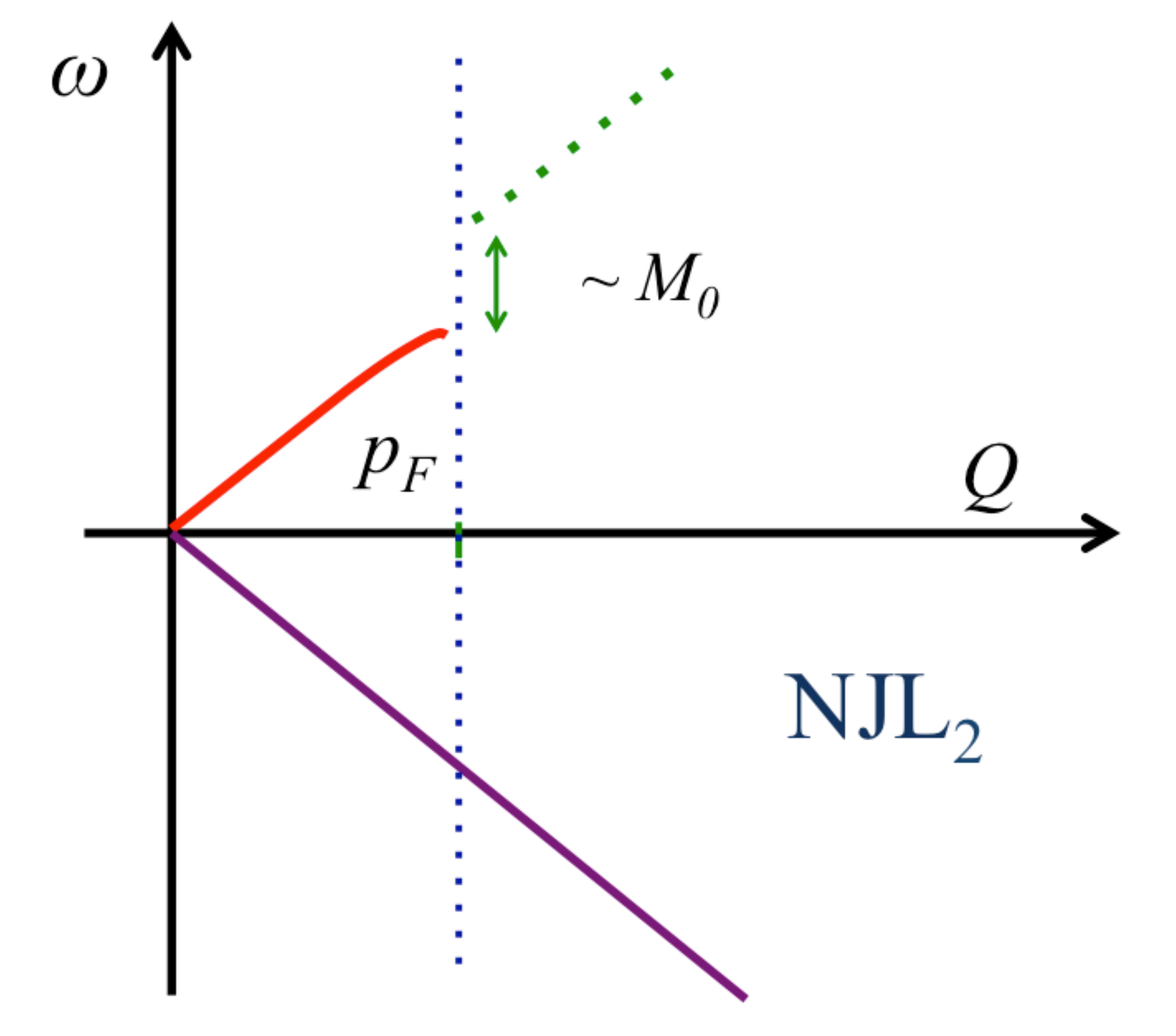} }
\end{center}
\vspace{-0.4cm}
\caption{The quasimomentum-energy dispersion. 
Left: For the GN model.
The spectra contain the gapped region
in the Fermi and Dirac sea.
The size of the gap decreases like $\sim M_0\times M_0/p_F$
as $p_F$ becomes large.
Right: For the NJL$_2$-type models.
The gaps open only at the Fermi points.
The size of the gap is known to be $\sim M_0$,
independently of the value of $p_F$.
}
\label{fig:FermiDirac}
\vspace{0.1cm}
\end{figure}

In this section we examine
the qualitative differences between 
the chiral spirals in the GN model and
in the QCD$_2$ or NJL$_2$ models.
First, we compare results of the GN model and of QCD$_2$
by contrasting the short- and long-range interactions.
Secondly, we argue why results of the GN and NJL$_2$ models
are qualitatively different,
by emphasizing the structure of the 4-Fermi interactions.

\subsection{Short-range versus long-range interactions}
\label{}

First let us recall the structure of 
the single-particle energy levels in the GN model.
The energy level has gaps in the Fermi and Dirac seas;
see Fig. \ref{fig:FermiDirac}.
In the chiral scalar density,
contributions from 
the first energy branch in the Fermi and Dirac sea
cancel out, leaving only the contribution
from the second energy branch in the Dirac sea
[see Eq. (\ref{scalarintegral})],
\begin{equation}
\la \bar{\psi} \psi \ra_{ {\rm I} }^{ {\rm Fermi} }
+ \left(\, \la \bar{\psi} \psi \ra_{ {\rm I} }
+ \la \bar{\psi} \psi \ra_{ {\rm II} } \,\right)^{ {\rm Dirac} }
= \la \bar{\psi} \psi \ra_{ {\rm II} }^{ {\rm Dirac} } \,.
~~~~~~({\rm GN~model})
\end{equation}
The condensate includes
the contributions up to $\omega \sim - \Lambda$.
This is the reason why the scalar density 
is proportional to $\sim \ln (2\Lambda/M_0)$.

On the other hand, in the chiral pseudoscalar density,
the Dirac sea contributions from the first and second
energy branches cancel,
leaving only the Fermi sea contribution,
\begin{equation}
\la \bar{\psi} \rmi \gamma_5 \psi \ra_{ {\rm I} }^{ {\rm Fermi} }
+ \left(\, \la \bar{\psi} \rmi \gamma_5 \psi \ra_{ {\rm I} }
+ \la \bar{\psi} \rmi \gamma_5 \psi \ra_{ {\rm II} } \,\right)^{ {\rm Dirac} }
= \la \bar{\psi} \rmi \gamma_5 \psi \ra_{ {\rm I} }^{ {\rm Fermi} } 
~~~~({\rm GN~model})\,.
\end{equation}
The amplitude is proportional to $\sim \ln (4 p_F/M_0)$.
Due to the mismatch in the net contributions
for the scalar and pseudoscalar density,
their amplitudes are naturally different in the GN model.

When using this result as a guide for the QCD phase diagram,
espcially when $p_F$ becomes larger than the vacuum 
quark effective mass,
we should use the GN results with some caution.
The above result strongly depends on the fact
that the gaps in the Fermi and Dirac sea have the same size
at the edge of the first energy branches.
Such large gaps in the Dirac sea
are rather specific to models with the contact interactions.
In such models, although the condensation is initially driven by the
low-energy particle-hole pairs near the Fermi surface,
the created condensate affects spectra
all the way from the Fermi surface down to the
Dirac sea.
Then the resulting gapped fermions in the Dirac sea
also contribute to the condensate,
giving large feedback to the fermions near the Fermi surface.
Therefore there is a tight connection
between the structures of the Fermi sea and Dirac sea.

In contrast, for models of the long-range interactions such as QCD,
the physics near the Fermi surface does not
strongly affect the structure of the Dirac sea.
In fact, with momentum-dependent forces,
the gap functions in general
become momentum dependent.
If we had used models of long-range interactions
such as $1/\vp^2$ force,
the gap would be large near the Fermi surface
but small otherwise.
In particular, the chirality-violating effective mass tends to
disappear in the Dirac sea as fermion density 
becomes large \cite{Kojo:2011cn}.
Then the main contribution to both
chiral scalar and pseudoscalar density comes from the
Fermi sea, and they tend to acquire the same amplitude.
Actually, this is what happens in models like QCD$_2$.

\subsection{The GN model versus the NJL$_2$ model}
\label{}

In the NJL$_2$ model,
the interaction is short range, as in the GN model.
Nevertheless, qualtiative aspects of the chiral condensates
are more similar to QCD$_2$ rather than the GN model.
Moreover, in contrast to the GN model,
models in the latter class have an energy gap
of $\sim M_0$ instead of a decreasing gap at finite density,
$\sim M_0^2/p_F$
(Fig. \ref{fig:FermiDirac}).
The key observation to understanding  all these tendencies is
that for a particular set of 4-Fermi interactions, 
physics near the Fermi surface
tends to decouple from physics in the Dirac sea,
as it happens for models with long-range interactions.

To explain this, first we project
the fermion fields onto the right- and left-moving components,
\begin{equation}
\psi_{r,\, l} = \frac{\, 1\pm \gamma_0 \gamma_z \,}{2} \psi\,,
~~~~~~\gamma_5 = \gamma_0 \gamma_z\,.
\end{equation}
For free fermions, the field equation is given by
\begin{equation}
(p_0 - p_z ) \psi_r =0\,,~~~~~~~ (p_0 + p_z ) \psi_l =0\,,
\label{dispersion_rl}
\end{equation}
from which we observe that
the right components have positive energy for $p_z>0$
and negative energy for $p_z<0$.
The relation is opposite for the left components.

Now we express the 4-Fermi interactions 
in terms of left and right components.
For bookkeeping purposes, let 
\begin{equation}
\Phi(x) \equiv \bar{\psi}_l \psi_r (x) \,,
~~~~~~~
\Phi^*(x) \equiv \bar{\psi}_r \psi_l (x) \,.
\end{equation}
The Fourier transform of $\Phi$ is
$\left(\int_x \equiv \int \rmd^2 x\,, 
\int_q \equiv \int \frac{\rmd^2 q}{\,(2\pi)^2 \,}\right)$
\begin{equation}
\Phi(q) = \int_{\delta p} 
\bar{\psi}_l \left(-\frac{q}{2}+\delta p \right)\, 
\psi_r \left(\frac{q}{2}+\delta p \right) \,.
\end{equation}
Notice that for $q_z \simeq 2 p_F$,
both of the fields $\psi_r$ and $\psi_l$ at small $\delta p_z$
describe the fermion fields near the Fermi points.
If $q_z$ is very different from $2p_F$,
either of the fields $\psi_l$ or $\psi_r$ in $\Phi(q)$ must
be in the region far away from the Fermi surface,
costing more energy.
This is the reason why the homogeneous 
condensation at $q_z=0$ tends to disappear at finite density 
while instead the inhomogeneous condensate 
of $\la \Phi(q_z=2p_F)\ra$ develops
due to the condensed
particle-hole pairs near the Fermi surface.

Now we consider
the 4-Fermi interaction in the NJL$_2$.
It can be written as
\begin{equation}
\int_x \left[ (\bar{\psi} \psi )^2 
+ (\bar{\psi} \rmi \gamma_5 \psi  )^2 \right]
= 4\int_x |\Phi(x)|^2 
= 4\int_q \Phi(q) \Phi^*(q) \,,
\end{equation}
Note that the interaction couples $\Phi(q)$ with
$\Phi^*(q)$.
When one $\Phi(q)$ is replaced with the mean field
$\la \Phi(q_z = 2p_F) \ra$,
it affects only $\Phi^*(q_z = 2p_F)$
in which left- and right-moving fermions can
simultaneously stay at low energy for small $\delta p_z$,
and also simultaneously go to high energy for large $\delta p_z$.
Phrasing this in another way,
the mean field scatters
low-energy fields to low energy,
and high-energy fields to high energy,
but it does not strongly mix up fields belonging to 
different energy domains.

The meaning of the above statements becomes
clearer if we consider the GN model. 
Its 4-Fermi interaction is
\begin{equation}
\int_x (\bar{\psi} \psi )^2 
=\! \int_x\! \left[\, 2|\Phi(x)|^2 + \Phi^2 + \Phi^{*2} \right]
=\! \int_q \left[\, 2\Phi(q) \Phi^*(q) 
+ \Phi(q) \Phi(-q) + \Phi^*(q) \Phi^*(-q) \right]\,,
\end{equation}
where we find extra couplings,
$\Phi(q) \Phi(-q)$ and $\Phi^*(q) \Phi^*(-q)$.
Now imagine that we have a mean field,
$\la \Phi(q_z = 2p_F) \ra$.
Then it couples to the composite field $\Phi(q_z = - 2p_F)$.
But its content is
\begin{equation}
\Phi(q_z = - 2p_F) = \int_{\delta p} 
\bar{\psi}_l \left(p_F+\delta p \right)\, 
\psi_r \left(-p_F+\delta p \right) \,.
\end{equation}
At small $\delta p_z$, both fields are in the Dirac sea
according to the dispersion (\ref{dispersion_rl}).
On the other hand, when
the right field $\psi_r$ stays near the Fermi surface,
then $\delta p_z \simeq 2p_F$,
and the $\psi_l$ field is in the Dirac sea 
with energy of $\sim -3p_F$.
Therefore, if the mean field is developed
by condensation near the Fermi surface,
it inevitably couples the fields near the Fermi surface
to those in the Dirac sea.
This is the reason why the GN model has
the sizable energy gap not only near the Fermi surface
but also in the Dirac sea.
Furthermore, the feedback from the Dirac sea condensation
strongly affects the Fermi surface condensations,
making the inhomogeneous solutions far more 
complicated than those in the NJL$_2$ model.
Because of this feedback,
the size of the mass gap in the GN model
scales like $\sim M_0^2/p_F$, unlike $\sim M_0$
in the NJL$_2$ model or QCD$_2$.

Perhaps it is already clear why the results of chiral condensates
in NJL$_2$ and QCD$_2$ are similar.
In the former, the combinations of the 4-Fermi interactions
are arranged in such a way that
fields for the Fermi surface do not strongly 
couple to those for the Dirac sea.
In the latter,
the long-range interactions
and the resulting momentum-dependent mass functions
tend to forbid
strong coupling between the Fermi surface
and Dirac sea.
In both cases, physics are governed by
dynamics near the Fermi surface.

\section{Summary}
\label{sec:summary}

In this paper, we have argued that
the inhomogeneous chiral condensate
in the GN model takes the chiral spiral form.
Although the thermodynamic functional 
does not depend on the pseudoscalar
density manifestly, the spatial modulations
of the chiral scalar density inevitably 
generate the pseudoscalar density modulations.

While our arguments add little  
for the understanding of the GN model,
the implications will become important
once we start to infer proper effective models
at finite quark density
from more general perspectives on the fundamental theories.
For assumed mean fields,
we should calculate not only the thermodynamic functional 
but also various operators
which might acquire large expectation values.
Once such operators are found,
we have to reanalyze the effective models
including the interaction terms
related to such operators,
unless there is some reason to discard them.
For the NJL$_4$ up to dimensions-6 operators,
it is perhaps unavoidable
to add $(\bar{\psi} \rmi \gamma_0 \gamma_j \psi)^2$-type 
4-Fermi interactions
whose two-dimensional counterpart is 
$(\bar{\psi} \rmi \gamma_5 \psi)_{ {\rm 2D} }^2$.
This is an approach pursued in Ref. \cite{Feng:2013tqa}.

We have also contrasted
the GN model with the NJL$_2$ model and QCD$_2$
to understand the structure of the chiral spirals
and the parametric behaviors of the mass gaps.
In the GN model,
the specific form of the 4-Fermi interaction 
and its short-range properties
together create a strong Fermi-Dirac sea coupling 
which deforms the Dirac sea, even producing the energy gap
inside the Dirac sea.
If the interaction is replaced with the long-range one,
such strong deformation of the Dirac sea
tends to disappear.
Similar results can be found 
if we arrange the 4-Fermi interactions
in such a way as to weaken couplings between the Fermi and Dirac seas, 
as happened in the NJL$_2$ model.

We found it very interesting that
simple arrangements of the 4-Fermi interactions
can control the coupling between the Fermi and Dirac
sea contributions.
Even within models with contact interactions,
we can reproduce results similar to those in models with 
long-range interactions.
Perhaps we may use the above kinematic considerations
as a guide to restrict possible forms 
of the effective models at finite density,
in addition to the ordinary 
symmetry considerations.

More implications from the GN model studies 
to the results of the NJL$_4$ model will be discussed elsewhere.

\section*{Acknowledgments}

T.K. acknowledges E.~J.~Ferrer, V.~de la Incera,
and S. Carignano
for very stimulating discussions which motivate this work
and for their kind hospitality during his visit to UTEP.
This research was supported in part by NSF Grants No. PHY09-69790 
and No. PHY13-05891.

\appendix

\section{Series expansion of the elliptic functions}

The series expansion for the elliptic functions
is useful for several purposes, such as numerical computations.
We have already given the expansion 
for Jacobi's $\theta$ function and $Z$ function
in Eqs. (\ref{thetaseries}) and (\ref{zetaseries}).
The expansions for the elliptic functions are
($q \equiv \rme^{-\pi \bfK'/\bfK}$) 
\footnote{Formula (16.23) in Ref. \cite{formula1}.}
\begin{align}
\sn(x|\lambda) 
&= \frac{2\pi}{\lambda^{1/2} \bfK}
\sum_{n=0} \frac{q^{n+1/2}}{\,1-q^{2n+1}\,} 
\sin \frac{\, (2n+1)\pi x \,}{2\bfK}\,,
\nonumber \\
\cn(x|\lambda) 
&= \frac{2\pi}{\, \lambda^{1/2} \bfK \,}
\sum_{n=0} \frac{q^{n+1/2}}{\, 1+q^{2n+1} \,} 
\cos \frac{\, (2n+1)\pi x \,}{2\bfK} \,,
\nonumber \\
\dn(x|\lambda) 
&= - \frac{\pi}{\, 2\bfK \,} 
+ \frac{2\pi}{\,\bfK \,}
\sum_{n=0} \frac{q^{n}}{\, 1+q^{2n} \,} 
\cos \frac{\, n\pi x \,}{\bfK} \,.
\label{sncndnseries}
\end{align}
For small $\lambda$,
the elliptic functions can be expanded as
\footnote{Formula (16.22) in Ref. \cite{formula1}.}
\begin{equation}
\sn(x|\lambda) = x - (1+\lambda) \frac{\, x^3 \,}{3!} + \cdots\,,
~~~
\cn(x|\lambda) = 1 - \frac{\, x^2 \,}{2!} + \cdots\,,
~~~
\dn(x|\lambda) = 1 - \lambda \frac{\, x^2 \,}{2!} + \cdots\,.
\label{smallxexpansion}
\end{equation}
%

\section{Relative phases}
\label{sec:relaphase}

To determine $\tchi$ from $\tvarphi$ including
the relative phase,
we use Eqs.(\ref{relation}) and (\ref{eq:phifix}).
We first compute the derivative.
First we note that
\begin{equation}
\frac{\rmd \tvarphi}{\rmd \xi}
= 
\left[\, \frac{\, \pi \,}{\, 2\bfK \,}
\left(\frac{1}{\, \theta_1 (u_{\xi+\alpha} ) \,}
\frac{\, \rmd \theta_1 (u_{\xi+\alpha}) \,}{\rmd u_{\xi+\alpha} }
- \frac{1}{\, \theta_4(u_{\xi} ) \,}
\frac{\, \rmd \theta_4(u_{\xi}) \,}{\rmd u_{\xi} } \right)
- Z(\alpha) \,\right]\tvarphi(\xi) \,.
\end{equation}
To proceed further, we need to use the formulas
\footnote{Formula (16.34) in Ref. \cite{formula1}.}.
\begin{equation}
\frac{1}{\, \theta_1 (u_a ) \,}
\frac{\, \rmd \theta_1 (u_a) \,}{\rmd u_a }
= \frac{\, 2\bfK \,}{\pi}
\left[\, Z(a) + \frac{\, \cn(a)\, \dn(a) \,}{\sn(a)} \right]\,,
~~~~~~~
\frac{1}{\, \theta_4 (u_a ) \,}
\frac{\, \rmd \theta_4 (u_a) \,}{\rmd u_a }
= \frac{\, 2\bfK \,}{\pi}\, Z(a) \,,
\end{equation}
with which we get
\begin{equation}
\frac{\, \rmd \tvarphi \,}{\rmd \xi}
= 
\left[\, Z(\xi+\alpha) 
+ \frac{\, \cn(\xi+\alpha)\, \dn(\xi+\alpha) \,}{\sn(\xi+\alpha)} 
- Z(\xi)
- Z(\alpha) \,\right] \tvarphi(\xi) \,.
\end{equation}
Next we use the addition theorem for the zeta function
\footnote{Formula (17.4.35) in Ref. \cite{formula1}.},
\begin{equation}
Z(\xi+\alpha) = Z(\xi) + Z(\alpha)
-\lambda\, \sn(\xi)\, \sn(\alpha) \, \sn(\xi+\alpha) \,,
\end{equation}
and for the elliptic functions
\footnote{Formula (16.17) in Ref. \cite{formula1}.},
\begin{align}
\sn(u+v) &
= \frac{\, \sn(u) \cdot \cn(v) \dn(v) + \sn(v) \cdot \cn(u) \dn(u) \,}
{1-\lambda \, \sn^2(u) \cdot \sn^2(v) } \,,
\nonumber \\
\cn(u+v) &
= \frac{\, \cn(u) \cdot \cn(v) - \sn(u) \dn(u) \cdot \sn(v) \dn(v) \,}
{1-\lambda \, \sn^2(u) \cdot \sn^2(v) } \,,
\nonumber \\
\dn(u+v) &
= \frac{\, \dn(u) \cdot \dn(v) -\lambda\, \sn(u) \cn(u) \cdot \sn(v) \cn(v) \,}
{1-\lambda \, \sn^2(u) \cdot \sn^2(v) } \,.
\end{align}
With these ingredients,
we have to do messy calculations and get
\begin{equation}
\tchi_\omega (\xi)
= - {\rm sgn}(\tomega) 
\frac{\,1 \,}{\, \dn(\xi) \,}\,
\frac{\, \cn(\xi+\alpha) \,}{\, \sn(\xi+\alpha) \,}\, 
\tvarphi_\omega (\xi)  \,.
\end{equation}
In the final step
we use the relations among the theta functions
and elliptic functions
[Sec.(22.11) in Ref. \cite{analysis}],
\begin{equation}
\sn(a) 
= \frac{\, \theta_3(0) \,}{\theta_2(0)}
\,\frac{\, \theta_1(u_a) \,}{\theta_4(u_a)} \,,
~~~~~
\cn(a) 
= \frac{\, \theta_4(0) \,}{\theta_2(0)}
\,\frac{\, \theta_2(u_a) \,}{\theta_4(u_a)} \,,
~~~~~
\dn(a) 
= \frac{\, \theta_4(0) \,}{\theta_3(0)}
\,\frac{\, \theta_3(u_a) \,}{\theta_4(u_a)} \,,
\label{eq:analysis}
\end{equation}
and then use the expression for $\tvarphi(\xi)$.
The result is
\begin{equation}
\tchi_\omega (\xi) 
= - {\rm sgn}(\tomega)\,
\frac{\, \theta_2 (u_{\xi+\alpha}) \,}{\, \theta_3 (u_{\xi}) \,}\, 
\frac{\, \theta_4 (u_{\xi}) \,}{\, \theta_1 (u_{\xi+\alpha}) \,}\, 
\tvarphi_\omega (\xi)
= - \calN {\rm sgn}(\tomega)\,
\frac{\, \theta_2 (u_{\xi+\alpha}) \,}{\, \theta_3 (u_{\xi}) \,}\, 
\rme^{\xi Z(\alpha) }
\,.
\end{equation}
This expression can be converted into a more convenient form.
To do this, we note that by definition
the functions $\theta_2$ and $\theta_3$
are related to $\theta_1$ and $\theta_4$ as
\begin{equation}
\theta_2(u) = \theta_1 (u+u_\bfK) = - \theta_1 (u-u_\bfK)
\,, ~~~~~ 
\theta_3(u) = \theta_4 (u+u_\bfK) = \theta_4 (u-u_\bfK) \,, 
\end{equation}
so the function 
$\tchi(\xi)$ is proportional to $\tvarphi(\xi-\bfK)$,
\begin{equation}
\tchi_\omega (\xi) 
= {\rm sgn}(\tomega)
 \, \rme^{\bfK Z(\alpha)}
\times \calN
\frac{\, \theta_1 (u_{\xi-\bfK +\alpha}) \,}{\, \theta_4 (u_{\xi-\bfK}) \,}\, 
\rme^{(\xi - \bfK)  Z(\alpha) }
= 
{\rm sgn}(\tomega)
\, \rme^{\bfK Z(\alpha)}
\times \tvarphi_\omega (\xi - \bfK)
\,,
\end{equation}
as it should be.
Note that the exponent is purely imaginary.

\section{Normalization factor}
\label{sec:normalization}

The equation to determine the normalization factor is
\begin{equation}
\frac{1}{\, 2\,}
= \frac{1}{\, 2\bfK\,}
\int_0^{2\bfK} \!\rmd \xi \, |\tvarphi_\omega (\xi)|^2
=\frac{\,  |\calN_\omega |^2 \,}{\, 2\bfK\,}\int_0^{2\bfK} \!\rmd \xi \,
\frac{\,\theta_1(u_{\xi+\alpha}) \theta_1(u_{\xi+\alpha^*})\,}
{\, \theta^2_4(u_{\xi})\,}\,,
\end{equation}
for each energy $\omega$.
We use the formula
\begin{equation}
\theta^2_3(0) \theta_1 (a+b) \theta_1(a-b)
= \theta_4^2(a) \theta_2^2(b) - \theta_4^2(b) \theta_2^2(a) \,.
\end{equation}
Similar formulas and derivations
can be found in Chap. 21 in Ref. \cite{analysis}.

(i) The $\alpha = \rmi \eta$ case.
In this case, $\alpha^* = -\alpha$,
and we get
\begin{equation}
\frac{\,\theta_1(u_{\xi+\alpha }) \theta_1(u_{\xi-\alpha})\,}
{\, \theta^2_4(u_{\xi})\,}
= \frac{\, \theta_2^2(u_{\alpha}) \,}{\, \theta_3^2(0) \,}
\left[\, 1
- \frac{\, \theta_4^2(u_{\alpha}) \,}{\, \theta_2^2(u_{\alpha} ) \,}
\frac{\, \theta_2^2(\xi) \,}{\, \theta_4^2(\xi) \,}
\, \right]
= \frac{\, \theta_2^2(u_{\alpha}) \,}{\, \theta_3^2(0) \,}
\left[\, 1
- \frac{\, \cn^2(\xi) \,}{\, \cn^2(\alpha) \,}
\, \right]
\,,
\label{eq:formula_eta}
\end{equation}
where we have used 
Eq.(\ref{eq:analysis}).
Finally, we take the spatial integral.
Recalling Eq.(\ref{eq:secondkind}),
we find
\begin{equation}
|\calN_\omega|^{-2}
= \frac{2}{\, \lambda \cn^2 (\alpha) \,}
 \frac{\, \theta_2^2(u_{\alpha}) \,}{\, \theta_3^2(0) \,}
\left[\, \dn^2(\alpha)
- \frac{\, \bfE \,}{\, \bfK \,}
\, \right]\,.
\label{eq:normal_eta}
\end{equation}
Combining Eqs.(\ref{eq:formula_eta})
and (\ref{eq:normal_eta}),
we can arrive at the expression
(\ref{normalized}).

(ii) The $\alpha =\bfK + \rmi \eta$ case.
We recall the relation (\ref{peritheta}),
$\theta_1(u_{\xi+2\bfK}) = - \theta_1(u_{\xi})$.
Then 
\begin{equation}
\theta_1(u_{\xi+\bfK+\rmi \eta}) \theta_1(u_{\xi+\bfK -\rmi \eta})
= - \theta_1(u_{\xi+\bfK+\rmi \eta}) \theta_1(u_{\xi-\bfK -\rmi \eta}) \,,
\end{equation}
so 
$\theta_1(u_{\xi+\alpha}) \theta_1(u_{\xi+\alpha^*})
=-\theta_1(u_{\xi+\alpha}) \theta_1(u_{\xi-\alpha})$.
Therefore we can obtain results 
corresponding to Eqs.(\ref{eq:formula_eta})
and (\ref{eq:normal_eta})
by multiplying them by $(-1)$.
On the other hand, the $(-1)$ factors
will cancel for the normalized probability function
$|\tvarphi(\xi)|^2$,
so Eq.(\ref{normalized})
takes the same form
for the $\alpha=\rmi \eta$ and $\bfK+\rmi \eta$ cases.

\section{Derivative of the Zeta-function}
\label{sec:derivativezeta}

To compute the derivative of the zeta function,
we use the expression
\footnote{Formula (17.4.28) in Ref. \cite{formula1}.}
\begin{equation}
Z(\alpha) = E(\alpha) - \alpha \bfE/\bfK \,,
\label{Z}
\end{equation}
where $E(\alpha)$ is the Jacobi incomplete
elliptic integral,
\begin{equation}
E (\alpha) 
= \int^{\alpha}_0 \!  \rmd w ~ \dn^2 w \,.
\end{equation}
In particular, for $\alpha = \bfK$, 
we have the complete integral, $E(\bfK) = \bfE$.
From these expressions we can easily 
arrive at Eq.(\ref{eq:zetaderivative}).

\section{More on the UV cutoff}
\label{sec:mapping}

Here we will complete the discussions outlined in Sec.\ref{sub:mu}.
First, we express $\omega_\Lambda$ by $\epsilon$.
Using the Jacobi imaginary transformation
and then applying the relation for the quarter period, we get
[remember that $\bfK'(\lambda) = \bfK(\lambda_1)$]
\begin{equation}
\frac{\, \omega_\Lambda \,}{\calA}
= \dn\left(\, \rmi(\bfK'-\epsilon) |\lambda \,\right)
= \frac{\, \dn(\bfK'-\epsilon\,|\lambda_1) \,}
{\, \cn(\bfK'-\epsilon\, |\lambda_1) \,}
= 
\frac{\, 1\,}{\, \sn(\epsilon\, |\lambda_1) \,} \,.
\end{equation}
Then, expanding the elliptic functions via Eq.(\ref{smallxexpansion}),
we get
\begin{equation}
\frac{\, \omega_\Lambda \,}{\calA}
= \frac{1}{\, \epsilon \,}
\left[\, 1 + (1+\lambda_1) \, 
\frac{\, \epsilon^2 \,}{3!} + \cdots \right]\,.
\label{omegaepsilon}
\end{equation}
Next, we treat our dispersion
at quasimomenum $Q=\Lambda$,
\begin{equation}
\frac{\, \Lambda \,}{\, \calA\,} 
= - \rmi  Z \left(\, \rmi(\bfK'-\epsilon)|\lambda \, \right) 
+ \frac{\, \pi \,}{\, 2\bfK \,} \,.
\end{equation}
Using the imaginary transformation formula
\footnote{Formula (17.4.36) in Ref. \cite{formula1}.},
\begin{equation}
\rmi Z \left(\, \rmi(\bfK'-\epsilon)|\lambda \, \right) 
= Z(\bfK'-\epsilon|\lambda_1) 
- \dn(\bfK'-\epsilon|\lambda_1) \,
\frac{\, \sn(\bfK'-\epsilon|\lambda_1) \,}
{\, \cn(\bfK'-\epsilon|\lambda_1) \,}
+ \frac{\pi}{\, 2\bfK \,} 
\left(1-\frac{\epsilon}{\bfK'} \right) \,.
\end{equation}
The computation of the first term on the rhs requires
comment.
Using Eq.(\ref{Z}) and then formulas
\footnote{Formulas (17.4.28) 
and (17.4.7) in Ref. \cite{formula1}.},
we get
\begin{align}
Z(\bfK'-\epsilon|\lambda_1)
&= E(\bfK'-\epsilon|\lambda_1) 
- \frac{\, \bfE' \,}{\bfK'} \left(\bfK'-\epsilon \right)
\nonumber \\
&= \bfE' - E(\epsilon|\lambda_1)
+ \lambda_1 
\frac{\, \sn(\epsilon|\lambda_1) \cn(\epsilon|\lambda_1)\,}
{\dn(\epsilon|\lambda_1)}
- \frac{\, \bfE' \,}{\bfK'} \left(\bfK'-\epsilon \right)
\nonumber \\
&= - \epsilon \left( \lambda - \frac{\, \bfE' \,}{\bfK'} \right)
 + O(\epsilon^3) \,.
\end{align}
The remaining calculations are straightforward.
We can express the momentum cutoff $\Lambda$
as a function of $\epsilon$,
\begin{equation}
\frac{\, \Lambda \,}{\, \calA\,} 
= 
\frac{1}{\, \epsilon \,}
+ \epsilon\, \left[\, 
\frac{\, \lambda -2\,}{3} 
+ \frac{\bfE}{\, \bfK \,} 
\,\right] \,.
\label{lambdaepsilon}
\end{equation}
Combining Eqs.(\ref{omegaepsilon}) and (\ref{lambdaepsilon})
to erase $\epsilon$,
we arrive at Eq.(\ref{momenergycutoff})
which expresses $\omega_\Lambda$ as a function of $\Lambda$.

\section{Location of the Fermi momentum}
\label{sec:location}

In the main text, we have assumed that
the location of the Fermi momentum coincides
with the momentum at which the gaps open. 
In the following, we will verify this statement.

Suppose that the Fermi energy is larger 
than the energy where the gaps open.
We write the Fermi energy as
\begin{equation}
\epsilon_F = \omega_F' + \delta \omega =\calA +\delta \omega\,,
~~~~~\delta \omega >0\,,
\end{equation}
where $\omega_F'=\calA$ is the minimum energy for the 
second energy branch.
With nonzero $\delta \omega$,
the energy density $\calE$ 
is a function of $(\lambda, \calA, \delta \omega)$.
We will verify that the energy density increases for 
$\delta \omega > 0$.

First, we use the number density constraint to write
$\calA$ as a function of $\lambda$ and $\delta \omega$.
Because the domain of the integration is changed by
the existence of $\delta \omega$,
we get the extra term in addition to terms we got before.
It is given by ($\tilde{\omega}=\omega/\calA$)
\begin{equation}
\int_\calA^{\calA + \delta \omega} 
 \! \frac{\, \rmd \omega \,}{\pi} \, 
\frac{\, \tomega^2 - \bfE/\bfK \, }
{\sqrt{(\tomega^2-1) (\tomega^2- \lambda_1 )\,} \,}
\simeq 
\frac{1}{\, \pi \,}
\left( 1 -\frac{ \bfE}{\bfK} \right) 
\sqrt{\frac{\, \calA \delta \omega \,}{\lambda} }  \,.
\end{equation}
Therefore, the number constraint is
\begin{equation}
\frac{\calA}{\, 2\bfK \,} 
+ \frac{1}{\, \pi \,}
\left( 1 -\frac{ \bfE}{\bfK} \right) 
\sqrt{\frac{\, \calA \delta \omega \,}{\lambda} }  + O(\delta \omega)
= \frac{\, p_F \,}{\pi} \,.
\end{equation}
This characterizes $\calA$ as a function of $\lambda$ and $\delta
\omega$, so $\calA = \calA(\lambda , \sqrt{\delta \omega} )$.
Now we expand $\calA$ with respect to $\sqrt{\delta \omega}$,
and write
\begin{equation}
\frac{\calA_0(\lambda)}{\, 2\bfK \,} = \frac{\, p_F \,}{\pi} \,,
~~~~~
\frac{\delta \calA(\lambda,\sqrt{\delta \omega})}{\, 2\bfK \,} = 
-\frac{1}{\, \pi \,} \left( 1 -\frac{ \bfE}{\bfK} \right) 
\sqrt{\frac{\, \calA_0\,}{\lambda} }  
\sqrt{ \delta \omega} ~~<0\,.
\label{deltaA}
\end{equation}
Here we have used the fact $\lambda_1 \le \bfE/\bfK \le 1$.
With this expression, we can write the energy density
as a function of two independent variables,
$\calE(\lambda, \sqrt{\delta \omega})$.

Like the computation for the number density,
the change of integration domain adds an extra term
to the energy density.
It is given by
\begin{equation}
\frac{\delta \calE}{N} =
\int_{\calA}^{\calA +\delta \omega} 
 \! \frac{\, \rmd \omega \,}{\pi} \, \calD(\omega) \, \omega
\simeq 
\frac{\, \calA \,}{\, \pi \,}
\left( 1 -\frac{ \bfE}{\bfK} \right) 
\sqrt{\frac{\, \calA \delta \omega \,}{\lambda} } 
\simeq
- \frac{\calA_0 \delta \calA}{\, 2\bfK \,} +O(\delta \omega)
\,,
\end{equation}
where we have used Eq.(\ref{deltaA}).
Then the total energy density (homogeneous part) is 
\begin{equation}
\frac{\,  \bar{\calE}^R\,}{N}
\simeq - \frac{\, \calA^2 \,}{\, 4\pi \,}
\left[\,  \left( 2-\lambda  - 2\, \frac{\, \bfE \,}{\bfK} \right) 
\ln \frac{\, M_0^2 \,}{\, \lambda \calA^2 \,} 
+ \left( 2-\lambda  - 4\, \frac{\, \bfE \,}{\bfK} \right) 
\, \right]
- \frac{\, \calA_0 \delta \calA \,}{\, 2 \bfK \,}
\,.
\end{equation}
Next, we expand the energy density around
$\delta \omega =0$ and
$\lambda=\lambda_0$ 
such that $M_0=\sqrt{\lambda_0} \calA_0(\lambda_0)$.
Because we verified that 
$\partial_\lambda \calE|_{\lambda =\lambda_0,  \delta \omega=0} =0$
in Eq.(\ref{minimization}),
the correction of $\delta \lambda=\lambda - \lambda_0$ 
starts at the quadratic order.
Therefore, the leading correction to $\calE(\lambda_0,\delta \omega=0)$
starts with the $\sqrt{\delta \omega}$ term, and is given by
\begin{equation}
\frac{\,  \delta \calE \,}{N}
=\frac{\, \calA_0 \delta \calA \,}{\, \pi \,}
\left[\, 
 \, \frac{\, \bfE \,}{\bfK} - \frac{\, \pi \,}{2} \,\right] 
~~ >0 ~~~~(\delta \calA<0\,,~~ \bfE/\bfK < \pi/2)
\,.
\end{equation}
This is the energy cost.
Therefore, to minimize the energy,
we must set $\delta \calA \propto \sqrt{\delta \omega}$ to zero.
We can repeat similar arguments for
a Fermi momentum smaller than the momentum at the gapped point.
This completes the proof.



\begin{thebibliography}{00}

\bibitem{Deryagin:1992rw}
For early attempts, see
  D.~V.~Deryagin, D.~Y.~Grigoriev, and V.~A.~Rubakov,
  Int.\ J.\ Mod.\ Phys.\ A {\bf 7} (1992) 659;
  E.~Shuster and D.~T.~Son,
  Nucl.\ Phys.\ B {\bf 573} (2000) 434
  [hep-ph/9905448];
  B.~-Y.~Park, M.~Rho, A.~Wirzba and I.~Zahed,
  Phys.\ Rev.\ D {\bf 62} (2000) 034015
  [hep-ph/9910347];
  R.~Rapp, E.~V.~Shuryak and I.~Zahed,
  Phys.\ Rev.\ D {\bf 63} (2001) 034008
  [hep-ph/0008207].

\bibitem{Buballa:2014tba}
For review,
  M.~Buballa and S.~Carignano,
  arXiv:1406.1367 [hep-ph].

\bibitem{Nickel:2009ke}
  D.~Nickel,
  Phys.\ Rev.\ Lett.\  {\bf 103} (2009) 072301
  [arXiv:0902.1778 [hep-ph]];
  {\it ibid.}
  Phys.\ Rev.\ D {\bf 80} (2009) 074025
  [arXiv:0906.5295 [hep-ph]];
  S.~Carignano, D.~Nickel and M.~Buballa,
  Phys.\ Rev.\ D {\bf 82} (2010) 054009
  [arXiv:1007.1397 [hep-ph]].

\bibitem{Abuki:2011pf}
  H.~Abuki, D.~Ishibashi and K.~Suzuki,
  Phys.\ Rev.\ D {\bf 85} (2012) 074002
  [arXiv:1109.1615 [hep-ph]];
  E.~Nakano and T.~Tatsumi,
  Phys.\ Rev.\ D {\bf 71} (2005) 114006
  [hep-ph/0411350].

\bibitem{Kojo:2009ha}
  T.~Kojo, Y.~Hidaka, L.~McLerran and R.~D.~Pisarski,
  Nucl.\ Phys.\ A {\bf 843} (2010) 37
  [arXiv:0912.3800 [hep-ph]].

\bibitem{Kojo:2011cn}
  T.~Kojo, Y.~Hidaka, K.~Fukushima, L.~D.~McLerran and R.~D.~Pisarski,
  Nucl.\ Phys.\ A {\bf 875} (2012) 94
  [arXiv:1107.2124 [hep-ph]];
  T.~Kojo, R.~D.~Pisarski and A.~M.~Tsvelik,
  Phys.\ Rev.\ D {\bf 82} (2010) 074015
  [arXiv:1007.0248 [hep-ph]].

\bibitem{Thies:2003br}
  M.~Thies,
  Phys.\ Rev.\ D {\bf 69} (2004) 067703
  [hep-th/0308164].
\bibitem{Thies:2003kk}
  M.~Thies and K.~Urlichs,
  Phys.\ Rev.\ D {\bf 67} (2003) 125015
  [hep-th/0302092].

\bibitem{Schon:2000he}
  V.~Schon and M.~Thies,
  Phys.\ Rev.\ D {\bf 62} (2000) 096002
  [hep-th/0003195];
  B.~Bringoltz,
  Phys.\ Rev.\ D {\bf 79} (2009) 105021
  [arXiv:0811.4141 [hep-lat]];
{\it ibid.}
{\bf 79} (2009) 125006
  [arXiv:0901.4035 [hep-lat]];
  T.~Kojo,
  Nucl.\ Phys.\ A {\bf 877} (2012) 70
  [arXiv:1106.2187 [hep-ph]].

\bibitem{Basar:2009fg}
  G.~Basar, G.~V.~Dunne and M.~Thies,
  Phys.\ Rev.\ D {\bf 79} (2009) 105012
  [arXiv:0903.1868 [hep-th]].

\bibitem{Feng:2013tqa}
  B.~Feng, E.~J.~Ferrer and V.~de la Incera,
  arXiv:1304.0256 [nucl-th].

\bibitem{formula1}
{\it Handbook of Mathematical Functions}, 
edited by M. Abramowitz and I. Stegun (Dover, New York, 1990).

\bibitem{analysis}
E. T. Whittaker and G. N. Watson, 
{\it A Course of Modern Analysis}, 
(Cambridge University Press, England, 1980).

\bibitem{Kusnezov}
H. Li, D. Kusnezov and F. Iachello, 
J. Phys. A: Math. Gen. {\bf 33}, 6413 (2000).


\bibitem{Dunne:1997ia}
  G.~V.~Dunne and J.~Feinberg,
  Phys.\ Rev.\ D {\bf 57} (1998) 1271
  [hep-th/9706012].

\end{thebibliography}
\end{document}